\documentclass[journal]{IEEEtran}
\usepackage{amsfonts}
\usepackage{amssymb}
\usepackage{amsmath}
\usepackage{algorithm}
\usepackage{algorithmic}
\usepackage{multirow}
\usepackage{xcolor}
\usepackage{graphicx}
\usepackage{cite}
\usepackage{exscale}
\usepackage{relsize}
\usepackage{subeqnarray}
\usepackage{cases}
\ifCLASSOPTIONcompsoc
\usepackage[caption=false,font=normalsize,labelfont=sf,textfont=sf]{subfig}
\else
\usepackage[caption=false,font=footnotesize]{subfig}
\fi
\DeclareMathOperator*{\argmax}{argmax}
\newcommand{\bm}[1]{\mbox{\boldmath{$#1$}}}

\newtheorem{Theo}{Theorem}
\newtheorem{Prop}{Proposition}

\usepackage[T1]{fontenc}

\begin{document}

\title{Sequential Bayesian Detection of Spike Activities from Fluorescence Observations}

\author{\IEEEauthorblockN{Zhuangkun~Wei,
                          Bin~Li, Weisi~Guo, Wenxiu Hu, and Chenglin~Zhao}

\thanks{Zhuangkun~Wei, Weisi Guo and Wenxiu Hu are with the School of Engineering, the University of Warwick,
West Midlands, CV47AL, UK. (Email: zhuangkun.wei@warwick.ac.uk).}
\thanks{Bin Li and Chenglin Zhao are with the School of Information and Communication Engineering (SICE),
Beijing University of Posts and Telecommunications (BUPT), Beijing,
100876, China.}
}

\maketitle

\begin{abstract}
Extracting and detecting spike activities from the fluorescence observations is an important step in understanding how neuron systems work. The main challenge lies in that the combination of the ambient noise with dynamic baseline fluctuation, often contaminates the observations, thereby deteriorating the reliability of spike detection. This may be even worse in the face of the nonlinear biological process, the coupling interactions between spikes and baseline, and the unknown critical parameters of an underlying physiological model, in which erroneous estimations of parameters will affect the detection of spikes causing further error propagation. In this paper, we propose a random finite set (RFS) based Bayesian approach. The dynamic behaviors of spike sequence, fluctuated baseline and unknown parameters are formulated as one RFS. This RFS state is capable of distinguishing the hidden active/silent states induced by spike and non-spike activities respectively, thereby \emph{negating the interaction role} played by spikes and other factors. Then, premised on the RFS states, a Bayesian inference scheme is designed to simultaneously estimate the model parameters, baseline, and crucial spike activities. Our results demonstrate that the proposed scheme can gain an extra $12\%$ detection accuracy in comparison with the state-of-the-art MLSpike method.
\end{abstract}


\IEEEpeerreviewmaketitle

\section{Introduction}
\IEEEPARstart{R}{ecent} advances in developing reliable and high resolution imaging and scanning devices, e.g., fluorescence microscopy \cite{6112225}, two-photon laser scanning microscopies \cite{Denk73, Zipfel2003Nonlinear} and acousto-optic (AO) random-access scanning \cite{Katona2012Fast}, have underpinned rapid progress in obtaining high-fidelity data on activities of interest. This in turn, has spurred researches in brain science, cognition, and cell communications. Detecting neuron actions (e.g. spike activities) is key to characterizing the dynamic firing behaviors of neuron cells, and serves as an important basis for understanding how neuron systems react and communicate as well as real brain functions. On the synthetic bio-engineering front, emerging research in Internet-of-Nano-Things (IoNT) has led to synthetic molecular communication devices that mimic neuron and cell signalling \cite{6708551}.

In the context of neural activity detection, spikes need to be detected through the fluorescence observations from a calcium image, which contains three levels of uncertainty. Firstly, spikes lead to a rapid rise in intracellular calcium ($\text{Ca}^{2+}$) concentration followed by a slow decay (i.e., time-to-peak $8$-$40$ms, and decay constant $0.3$-$1.5$s \cite{Grewe2010High, Chen2013Ultrasensitive, Dana2014Thy1}). This effect gives rise to the transients induced by individual spikes to overlap, and aggregating in a nonlinear manner \cite{Akerboom2012Optimization}. Secondly, observations are often contaminated by significant ambient noise sources, which when combined with the dynamically fluctuated baseline, exhibits a coupling behavior that leads to interactions between the inferences of spikes and baseline, causing additional false/miss detection \cite{Deneux2016Accurate, Vogelstein2009Spike, Henry2013Inference}. Thirdly, critical parameters (e.g., the unitary $\text{Ca}^{2+}$ fluorescence transient's amplitude $A$ and decay time $\tau$) are inhomogeneous across neurons and cortical areas. Absence of accurately deriving these parameters may deteriorate the detection reliability.

\subsection{Related Works}
Over the past decade, several signal processing algorithms have been developed for spike detection. Existing algorithms can essentially be categorized into two groups.

The first group resorts to \emph{template matching} \cite{Grewe2010High, Kerr2005From, Kerr2007Spatial, Greenberg2008Population, Ozden2008, Ranganathan2010Optical}, whereby spikes are detected by matching their fluorescence observations to an empirically designed template. Unfortunately, the vulnerability to non-ergodic baseline signal and model parameters, as well as the inaccurate derivation of the spike template, limits their usages. Due to the inherent coupling between fluctuated baseline and spikes, false spikes are inevitably involved into template extraction, thereby deteriorating the template accuracy and the detection reliability. For instance, the state-of-the-art template matching scheme, Peeling \cite{Grewe2010High} is theoretically important, but is unable to distinguish between a real spike and a baseline imitation, not to mention that the involved complexity needed to extract the template and parameters grows exponentially with the desired accuracy.

The second group is underpinned by stochastic and probabilistic analysis. Such algorithms leverage on well-constructed physiological models that specify the nonlinearities between fluorescence observations and the genetically encoded calcium indicator (GECI). For instance, \cite{8309355} advanced an estimator that can detect two adjacent signals with inter-symbol-interference (ISI) on the order of tens of milliseconds. Unfortunately, this method overlooked the fluctuated baseline and assumed a known quantity of model parameters, therefore becoming less attractive to address the dynamic baseline issue. \cite{Vogelstein2009Spike, 6810293, Pnevmatikakis2016Simultaneous, Henry2013Inference, Deneux2016Accurate} proposed various inference schemes aiming at detecting the spikes as well as the baseline and model parameters. Specially, the state-of-the-art MLSpike \cite{Deneux2016Accurate} uses a maximum-likelihood method to derive an optimal spike train, by integrating an auto-calibration algorithm to statically estimate critical model parameters. However, these schemes overlook the complicated coupling and nonlinear interactions between spikes, baseline and model parameters. In other words, parameter calibration is in essence implemented in a static manner, unable to refine the estimated parameters as the detection process advances. In addition, without distinguishing whether the parameters are related with spike or background activities (referred as spike/background -related parameters), fixed calibration method in \cite{Deneux2016Accurate} may derive spike-related parameters via background signals. As a result, these inaccurate parameter estimations (e.g., up to a $20\%$ relative error of transient's amplitude $A$, and $30\%$ error of delay constant $\tau$ from MLSpike) cause catastrophic inference errors of the baseline and subsequent spike detection (about an error of $10\%$). This in turn will lead to propagating errors along the time-series. Furthermore, by enumerating all the joint assembles of discrete baseline and calcium concentration level, MLSpike also suffers a severe computational burden and thereby becomes less attractive for large data analysis.

\subsection{Contributions}
In this work, we suggest a sequential Bayesian estimation scheme based on a novel formation of random finite set (RFS), aiming to simultaneously detect and estimate potential spikes and fluctuated baseline, as well as consistently refine the unknown model parameters. To sum up, the main contributions of this paper are listed as follows:

(1) We formulate a RFS state to characterize the random spikes, fluctuated baseline, and unknown model parameters, so that they can be simultaneously estimated and refined in a dynamic way. In this way, errors induced by the stationary derivations of model parameters are alleviated. Moreover, the interaction between spikes and baseline is now uncoupled by the RFS state, which is capable of distinguishing the hidden active/silent states (induced by spike activities/non-spike). As such, spikes and their related parameters (e.g., fluorescence transient's amplitude $A$ and decay time $\tau$) are only estimated at the same time of detecting active spike trains, while the background parameters are estimated in the regions with both spikes and non-spike. Therefore, by inferring on the formulated RFS state, we are able to address both the coupling interactions and the erroneous parameter issues.

(2) To derive the RFS state, we design a sequential RFS-Bayesian inference scheme. Premised on the RFS concept, the proposed two-stage Bayesian scheme infers the RFS state by maximizing its posterior finite set density. Whilst the probability density function (PDF) is able to characterize the stochastic property of a random variable, the finite set statistics (FISST) PDF can simultaneously describe its existence and statistics. By presenting a two-stage recursive algorithm, this finite set posterior density is capable of characterizing both the active/silent status and the statistics of the RFS state. This allows us to \emph{decouple the spike activity sequence}.

(3) We evaluate the detection/estimation performance of the proposed RFS-Bayesian scheme from the noisy imaging data. Simulation results demonstrate the reliability of spike detection, and show that the accuracy of estimations of baseline and model parameters is significantly improved when compared to the state-of-the-art MLSpike algorithm. Also, the computational complexity of the proposed scheme is much lower than that of MLSpike, suggesting an advantage in computations. By casting accurate inferences of spikes, baseline and model parameters into an energy-efficient framework, the proposed RFS-Bayesian scheme presents great promises to not only the neural spike scenarios, but also other detection tasks in complex and dynamic scenarios.

The rest of this article is structured as follows. In Section II, we introduce the popular physiological model that characterizes the three levels of
uncertainties involved in neuron signalling dynamics. In Section III, the RFS-Bayesian scheme, including how to formulate the suitable RFS state and the new design of the two-stage Bayesian inference, is articulated. Numerical simulations are provided in Section IV. Finally, we conclude this study in Section V along with discussions of application areas and future impact.

\section{Model and Problem Formulation}

\begin{figure*}[!t]
\centering
\includegraphics[width=5.6in]{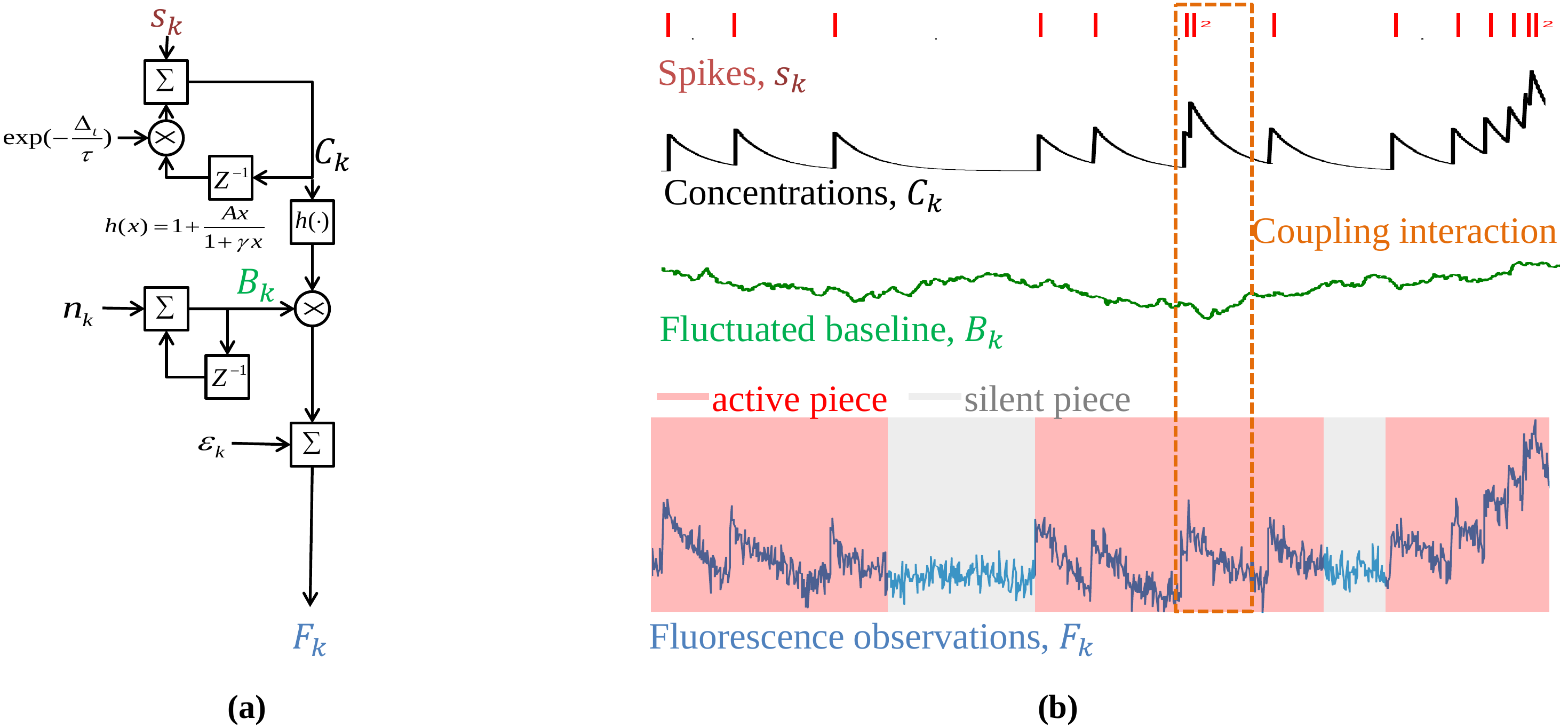}
\caption{System model for the involved physiological process and the illustration of calcium fluorescence. A schematic system structure is shown in (a), accompanying with the informative spikes, calcium concentrations, fluctuated baselines and the measured fluorescence observations in (b).}
\end{figure*}

\subsection{Physiological Model}
A physiological model that characterizes the evolution of calcium concentration $[\text{Ca}^{2+}]$ induced by the spike activities, dynamic baseline fluctuation, and fluorescence observations is given as follows \cite{Deneux2016Accurate, Vogelstein2009Spike, Henry2013Inference},
\begin{equation}
\frac{d[\text{Ca}^{2+}]}{dt}=\frac{-\gamma_e([\text{Ca}^{2+}]- [\text{Ca}^{2+}]_{\text{rest}})}{1+\kappa_S+\kappa_B}+\Delta[\text{Ca}^{2+}]_T\cdot s(t),
\end{equation}
\begin{equation}
F(t)-B(t)=(F_\text{max}-B(t))\cdot\frac{[\text{Ca}^{2+}]-[\text{Ca}^{2+}]_{\text{rest}}} {[\text{Ca}^{2+}]+K_d}+\varepsilon(t).
\end{equation}

Eq. (1) specifies the evolution of calcium concentration with time $t$. $s(t)$ is the binary spike activity, following a Poisson statistics with mean rate $\lambda$; $\Delta[\text{Ca}^{2+}]_T$ is the total calcium intracellular increase driven by one spike activity. $\Delta[\text{Ca}^{2+}]_T\cdot s(t))$ therefore gives the increment component of calcium concentrations. $\gamma_e$ denotes the calcium extrusion rate; $[\text{Ca}^{2+}]_{\text{rest}}$ is the free-calcium concentration at rest. So $\gamma_e([\text{Ca}^{2+}]- [\text{Ca}^{2+}]_{\text{rest}})$ describes the extrusion component of calcium. $\kappa_S$ and $\kappa_B$ are constant calcium binding ratios respectively of endogenous buffers and of the dye, which, combining with the aforementioned components, specifies the change of calcium concentration with respect to time $t$.

Eq. (2) provides the fluorescence observations. $F_0$ and $F_{\text{max}}$ are the fluorescence levels at rest and when the dye is saturated, respectively. $B(t)\in(F_0,~F_{\text{max}})$ gives the multiplicative fluctuated baseline, governed by the Brownian motion that is driven by process noise $n(t)$. $K_d$ is the dissociation constant of the dye. $\varepsilon(t)$ denotes the ambient noise. In this way, the fluorescence observations can be intuitively interpreted as the normalized and instantaneous response of current calcium concentration, plus fluctuated baseline and ambient noise.

Then by re-parameterizing Eqs. (1)-(2), a normalized intracellular calcium concentration is reformulated as:
\begin{equation}
C(t)\triangleq\frac{[\text{Ca}^{2+}]-[\text{Ca}^{2+}]_{\text{rest}}}{\Delta[\text{Ca}^{2+}]_T},
\end{equation}
a time-delay constant parameter:
\begin{equation}
\tau\triangleq\frac{1+\kappa_S+\kappa_B}{\gamma_e},
\end{equation}
a saturation parameter:
\begin{equation}
\gamma\triangleq\frac{\Delta[\text{Ca}^{2+}]_T}{[\text{Ca}^{2+}]_{\text{rest}}+K_d},
\end{equation}
and a transient amplitude:
\begin{equation}
A(t)\triangleq\frac{F_{\text{max}}-B(t)}{B(t)}\cdot\gamma.
\end{equation}
Note that given the limited variation range of $B(t)\in(F_0,~F_\text{max})$, $A(t)$ can be simplified as an unknown constant parameter $A$, with a known range \cite{Deneux2016Accurate}, i.e., $A\in[A_\text{min},~A_\text{max}]$.

By taking (3)-(6) into (1)-(2), and discretizing them by sample interval $\Delta_t$, the popular discrete physiological model is provided as follows \cite{Deneux2016Accurate}:
\begin{eqnarray}
C_k&=&\exp(-\frac{\Delta_t}{\tau})\cdot C_{k-1}+s_k\\
B_k&=&B_{k-1}+n_k\\
F_k&=&B_k\cdot(1+A\cdot\frac{C_k}{1+\gamma\cdot C_k})+\varepsilon_k
\end{eqnarray}
where Eqs. (7)-(8) are referred as dynamic functions, and Eq. (9) accounts for the fluorescence measurement function. The schematic flow of the neuron spiking system is illustrated in Fig. 1-(a).

i) The first dynamic function gives the evolution of the calcium concentrations (i.e., $C_k$), involving the number of spikes during the interval of $t_k-t_{k-1}$, denoted as $s_k\in\mathbb{N}$, and the partially unknown decay parameter $\tau$ (i.e., we may know in practice a coarse range of $\tau$ as $\tau_{\text{min}}<\tau<\tau_{\text{max}}$ \cite{Deneux2016Accurate}).

ii) The second dynamic function gives the evolution of the fluctuated baseline, which follows the 1-order Markov process, driven by the Gaussian process i.e., $n_k\sim\mathcal{N}(0,\eta^2)$. Note that, here $\eta^2$ is unknown to the receptor, which challenges the estimation process of the baseline.

iii) The measurement function describes the nonlinearity between concentrations (i.e., $C_k$), fluctuated baseline(i.e., $B_k$), and fluorescence observations (i.e., $F_k$). In common practice, the saturation parameter $\gamma$ is not a sensitive value for the measurement function, thus an average version is employed \cite{Deneux2016Accurate}. The ambient noise in Eq. (9) is assumed to be independent and identically distributed (i.i.d), Gaussian random variable, i.e., $\varepsilon_k\sim\mathcal{N}(0,\sigma^2)$, with an unknown variance.

\subsection{Problem Formulation}
Given the widely-accepted physiological model in Eqs. (7)-(9), the objective of this paper is thereby to simultaneously infer the concentration $C_k$ and baseline $B_k$ from a set of received fluorescence observation vectors, i.e., $\mathbf{F}_{1:k}\triangleq[F_1,F_2,...,F_k]^T$, under unknown parameters $\tau$, $\eta$, $A$ and $\sigma$.

The difficulties are emphasized as follows. Firstly, the observations suffer the coupling interactions between random spikes $s_k$ and the baseline. For instance, as shown in Fig. 1-(b), the two-spike induced observations are compromised by a series of small baseline, making them resemble those generated by one-spike. Secondly, the estimations of model parameters may be erroneous if we use background activities (e.g., the false spikes produced by dynamic transition of baseline) to acquire the spike-related parameters. Thirdly, given the nonlinearity between observations and the spike induced concentrations, the erroneous derivation of parameters affects severely the inference of spikes and baseline, and vice versa.

\section{RFS-Bayesian Scheme}
In order to cope with the difficulties, we propose the RFS-Bayesian scheme, which is capable of distinguishing the active/silent pieces for spike/background activities, and simultaneously inferring the concentration $C_k$, baseline $B_k$, and the model parameters $\tau$, $\eta$, $A$, and $\sigma$.

To do so, we firstly construct the RFS state of $C_k$, $B_k$, $\tau$, $\eta$, $A$, and $\sigma$ to address the coupling interactions, and to detect the active/silent pieces. Then, the two-stage Bayesian process, premised on the RFS state is designed, aiming to sequentially estimate the requested concentration, baseline and parameters.

\subsection{Random Finite Set Channel State Information}
A RFS, denoted as $\bm{\mathcal{U}}$, is one random set, whose cardinality (i.e. the number of elements in $\bm{\mathcal{U}}$, denoted as
$|\bm{\mathcal{U}}|$) and elements are all random variables varying with time. Analogous to the PDF in describing random variables, a RFS
$\bm{\mathcal{U}}=\{\mathbf{u}_1,\mathbf{u}_2,...,\mathbf{u}_M\}$ can be characterized completely via its cardinal density, $\varrho(M)$, and the joint distribution of its elements, $p(\mathbf{u}_1,\mathbf{u}_2,...,\mathbf{u}_M)$, i.e.,
\begin{equation}
f(\bm{\mathcal{U}})=M!\cdot\varrho(M)\cdot p(\mathbf{u}_1,\mathbf{u}_2,...,\mathbf{u}_M),
\end{equation}
which is referred to as the FISST PDF \cite{6497685,Mahler2007Statistical}.

In the context of spike detection, we construct the RFS state, denoted as $\bm{\mathcal{U}}_k$, for $C_k$, $B_k$, $\tau$, $\eta$, $A$, and $\sigma$, as follows:
\begin{equation}
\bm{\mathcal{U}}_k\subseteqq\Big\{\underbrace{[C_k,\tau,A]^T}_{\mathbf{u}_k}, ~\underbrace{[B_k,\eta,\sigma]^T}_{\mathbf{v}_k}\Big\},
\end{equation}
where vector $\mathbf{u}_k$ and $\mathbf{v}_k$ are two elements corresponding to spikes and background activities, respectively. As such, Eq. (11) has an interpretation as that, $\mathbf{u}_k$ appears and thus can be estimated only within the active pieces, while $\mathbf{v}_k$ never disappears from the observations and can be inferred in both the active/silent pieces. In this way, the RFS state indeed distinguishes the active/silent signals, as shown in Fig.1-(b).

\subsubsection{Cardinality of RFS State}
Given the definition of $\bm{\mathcal{U}}_k$ in Eq. (11), the cardinality of $\bm{\mathcal{U}}_k$ is either $1$ (i.e., $\bm{\mathcal{U}}_k=\{\mathbf{v}_k\}$) or $2$ (i.e., $\bm{\mathcal{U}}_k=\{\mathbf{u}_k,\mathbf{v}_k\}$). The probabilistic density of the cardinality is computed as follows:
\begin{equation}
\varrho(|\bm{\mathcal{U}}_k|)=
\begin{cases}
1-q_k & |\bm{\mathcal{U}}_k|=1,\\
q_k & |\bm{\mathcal{U}}_k|=2,\\
0 & \text{otherwise},
\end{cases}
\end{equation}
with a probability $q_k$ indicating whether it is silent or active in the time-slot $k$.

\subsubsection{FISST PDF of RFS State}
With the help of the cardinality in Eq. (12), the FISST PDF of $\bm{\mathcal{U}}_k$ is expressed by multiplying its cardinality density and the joint PDFs of its elements:
\begin{equation}
f(\bm{\mathcal{U}}_k)=
\begin{cases}
(1-q_k)\cdot p(\mathbf{v}_k) & \bm{\mathcal{U}}_k=\{\mathbf{v}_k\},\\
q_k\cdot p(\mathbf{u}_k,\mathbf{v}_k) & \bm{\mathcal{U}}_k=\{\mathbf{u}_k,\mathbf{v}_k\},\\
0 & \text{otherwise},
\end{cases}
\end{equation}
where $p(\cdot)$ denotes the (joint) PDFs of elements.

Then, inferring on the RFS state $\bm{\mathcal{U}}_k$ is thereby to compute its posterior density, and we need firstly study its transitional probabilities and likelihoods.

\subsubsection{Transitional FISST PDF of RFS State}
We use a hidden Markov process to describe the RFS state $\bm{\mathcal{U}}_k$. As shown in Fig. 2-(a), two hidden states (i.e., $|\bm{\mathcal{U}}_k|=1$ and $|\bm{\mathcal{U}}_k|=2$) specify respectively the silent and active pieces. Their transitional probabilities are denoted by the corresponding transition probability matrix (TPM):
\begin{equation}
\mathbf{\Pi}=\begin{bmatrix}
     1-P_b,~ P_b\\[3mm]
     1-P_s(\mathbf{u}_k),~ P_s(\mathbf{u}_k)\\
\end{bmatrix},
\end{equation}
where
$P_b\triangleq\text{Pr}\{|\bm{\mathcal{U}}_k|=2\big||\bm{\mathcal{U}}_{k-1}|=1\}\equiv 1-\exp(-\lambda)$ is defined as the silent-to-active probability, driven by a new arrival spike (seen from Fig. 2-(b)). $P_s(\mathbf{u}_k)\triangleq\text{Pr}\{|\bm{\mathcal{U}}_k|=2\big||\bm{\mathcal{U}}_{k-1}|=2\}$ is defined as the active-to-active probability.

It is noteworthy that, unlike the usual hidden states where transitional probabilities are constants, in this model, the active-to-active probability $P_s(\mathbf{u}_k)$ varies with the value of concentrations $C_k$, i.e.,
\begin{equation}
P_s(\mathbf{u}_k)=
{\begin{cases}
1 & C_k>C_\text{th},\\
0 & C_k\leq C_\text{th}.\\
\end{cases}}
\end{equation}
Here $C_\text{th}$ denotes the active/silent threshold, which is used to differentiate active and silent pieces. See from Fig. 2-(c). If $C_k>C_\text{th}$, $\bm{\mathcal{U}}_k$ belongs to the active piece, otherwise, $\bm{\mathcal{U}}_k$ falls into the silent piece. It is also noteworthy that the proposal of $C_\text{th}$ should be neither too large nor too small. This is because a large $C_\text{th}$ narrows the active region, which may lead to inadequate estimations of the spike-related model parameters; while a small $C_\text{th}$ broadens the active pieces, in which the spike induced concentration is almost extruded (i.e., $C_k\rightarrow0$), giving rise to ineffective inferences of those spike-related parameters.

\begin{figure}[!t]
\centering
\includegraphics[width=2.5in]{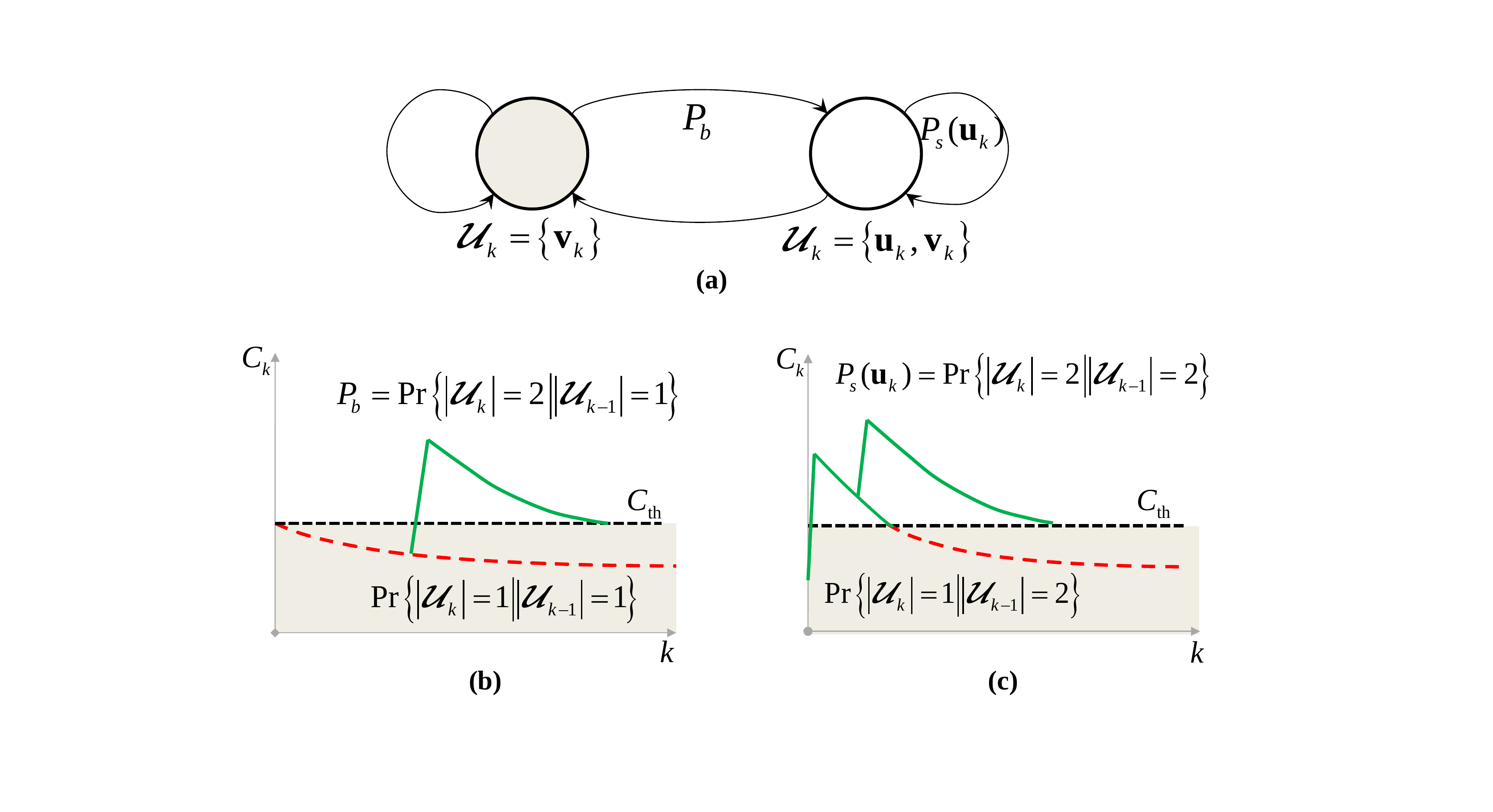}
\caption{Illustration of the hidden state Markov process of the RFS state.}
\end{figure}

Given the TPM of the hidden states, and the physiological model that involves both the transitions of concentration $C_k$ and baseline $B_k$, we compute the transitional FISST PDF of $\bm{\mathcal{U}}_k$, by taking possible transitions among hidden states and the associated parameter densities of $\mathbf{u}_k$ and $\mathbf{v}_k$, i.e.,
\begin{equation}
\begin{aligned}
&\phi_{k|k-1}(\bm{\mathcal{U}}_k|\{\mathbf{v}_{k-1}\})\\[3mm]
&=
{\begin{cases}
(1-P_b)\cdot\pi_{k|k-1}(\mathbf{v}_k|\mathbf{v}_{k-1}) & \bm{\mathcal{U}}_k=\{\mathbf{v}_k\},\\
P_b\cdot\pi_{k|k-1}(\mathbf{v}_k|\mathbf{v}_{k-1})\cdot b_k(\mathbf{u}_k) & \bm{\mathcal{U}}_k=\{\mathbf{u}_k,\mathbf{v}_k\},\\
0 & \text{otherwise},\\
\end{cases}}
\end{aligned}
\end{equation}
\begin{equation}
\begin{aligned}
&\phi_{k|k-1}(\bm{\mathcal{U}}_k|\{\mathbf{u}_{k-1},\mathbf{v}_{k-1}\})\\[3mm]
&=
{\begin{cases}
(1-P_s(\mathbf{u}_k))\cdot\pi_{k|k-1}(\mathbf{v}_k|\mathbf{v}_{k-1}) & \bm{\mathcal{U}}_k=\{\mathbf{v}_k\},\\
P_s(\mathbf{u}_k)\cdot\psi_{k|k-1}(\mathbf{u}_k,\mathbf{v}_k|\mathbf{u}_{k-1},\mathbf{v}_{k-1}) & \bm{\mathcal{U}}_k=\{\mathbf{u}_k,\mathbf{v}_k\},\\
0 & \text{otherwise},\\
\end{cases}}
\end{aligned}
\end{equation}
where $b_k(\mathbf{u}_k)$ is the activated density of $\mathbf{u}_k$ when new spikes burst. $\pi_{k|k-1}(\mathbf{v}_k|\mathbf{v}_{k-1})$, $\varpi_{k|k-1}(\mathbf{u}_k|\mathbf{u}_{k-1})$ and $\psi_{k|k-1}(\mathbf{u}_k,\mathbf{v}_k|\mathbf{u}_{k-1},\mathbf{v}_{k-1})$ denote the transitional PDFs, which can be computed by Eqs. (18)-(20) given that model parameters $A$, $\tau$, $\eta$, and $\sigma$ are constants.
\begin{equation}
\pi_{k|k-1}(\mathbf{v}_k|\mathbf{v}_{k-1})=p(B_k|B_{k-1})=\mathcal{N}(B_k; B_{k-1},\eta^2),
\end{equation}
\begin{equation}
\varpi_{k|k-1}(\mathbf{u}_k|\mathbf{u}_{k-1})=p(C_k|C_{k-1})= \frac{(\lambda\Delta_t)^{s_k}}{s_k!}\exp(-\lambda\Delta_t),
\end{equation}
\begin{equation}
\begin{aligned}
\psi_{k|k-1}(\mathbf{u}_k,&\mathbf{v}_k|\mathbf{u}_{k-1},\mathbf{v}_{k-1})\\ &=\pi_{k|k-1}(\mathbf{v}_k|\mathbf{v}_{k-1}) \cdot\varpi_{k|k-1}(\mathbf{u}_k|\mathbf{u}_{k-1}).
\end{aligned}
\end{equation}

The interpretations of transitional FISST PDFs in Eqs. (16)-(17) are given as follows. In the case of a silent piece at time-slot $k-1$ (i.e. $\bm{\mathcal{U}}_{k-1}=\{\mathbf{v}_{k-1}\}$), if the current time-slot $k$ is still occupied by silence (i.e., $\bm{\mathcal{U}}_k=\{\mathbf{v}_k\}$), the transitional density is composed only by the probability of silent-to-silent (i.e., $1-P_b$), and the transitional PDF from $\mathbf{v}_{k-1}$ to $\mathbf{v}_k$ (i.e., $\pi_{k|k-1}(\mathbf{v}_k|\mathbf{v}_{k-1})$). On the other hand, if one spike arrives at time-slot $k$ and results in $\bm{\mathcal{U}}_k=\{\mathbf{u}_k,\mathbf{v}_k\}$, the transitional density should contain information of both transitional $\pi_{k|k-1}(\mathbf{v}_k|\mathbf{v}_{k-1})$, and the activated density $b_k(\mathbf{u}_k)$, weighted by the silent-to-active cardinality probability $P_b$.

Considers the other case of an active signal at time-slot $k-1$ where $\bm{\mathcal{U}}_{k-1}=\{\mathbf{u}_{k-1},\mathbf{v}_{k-1}\}$. In this case, if time-slot $k$ is a silent piece (i.e., $\bm{\mathcal{U}}_k=\{\mathbf{v}_k\}$), then only the transitional PDF $\pi_{k|k-1}(\mathbf{v}_k|\mathbf{v}_{k-1})$, weighted by active-to-silent cardinality $1-P_s(\mathbf{u}_k)$, remains in the FISST PDF. And otherwise, the FISST PDF should involve the transitional density from both $\mathbf{u}_{k-1},\mathbf{v}_{k-1}$ to $\mathbf{u}_k,\mathbf{v}_k$, i.e., $\psi_{k|k-1}(\mathbf{u}_k,\mathbf{v}_k|\mathbf{u}_{k-1},\mathbf{v}_{k-1})$ weighted by the active-to-active probability $P_s(\mathbf{u}_k)$.

\subsubsection{Likelihood PDF conditioned on RFS State}
Provided the fluorescence observation $F_k$ of the current time-slot $k$, we can compute a likelihood PDF for the RFS state $\bm{\mathcal{U}}_k$, denoted as $\varphi_k(F_k|\bm{\mathcal{U}}_k)$, which is vital in further developing the two-stage Bayesian algorithm.

Given the formulation of $\bm{\mathcal{U}}_k$ in Eq. (11) as well as the dynamically physiological model in Eqs. (7)-(9), we will respectively compute the likelihoods as $\varphi_k(F_k|\{\mathbf{v}_k\})$, and $\varphi_k(F_k|\{\mathbf{u}_k,\mathbf{v}_k\})$.

In the case of $\bm{\mathcal{U}}_k=\{\mathbf{v}_k\}$, the observed fluorescence $F_k$ reflects the information corresponding with $\mathbf{v}_k=[B_k,\eta,\sigma]^T$. Thus, fluorescence $F_k$ in Eq. (9) can be re-written as $F_k=B_k+\varepsilon_k$, and the likelihood can be thereby computed as:
\begin{equation}
\varphi_k(F_k|\{\mathbf{v}_k\})=\mathcal{N}(F_k; B_k,\sigma^2).
\end{equation}

When $\bm{\mathcal{U}}_k=\{\mathbf{u}_k,\mathbf{v}_k\}$, fluorescence $F_k$ involves the whole $[C_k, \tau, A, B_k,\eta,\sigma]^T$. In this case, the likelihood PDF is determined as follows:
\begin{equation}
\varphi_k(F_k|\{\mathbf{u}_k,\mathbf{v}_k\})=\mathcal{N} \Big(F_k; B_k(1+A\frac{C_k}{1+\gamma C_k}),\sigma^2\Big).
\end{equation}

\subsubsection{Estimate of RFS State}
We denote a posterior FISST PDF of $\bm{\mathcal{U}}_k$ conditioned on the fluorescence observation vector $\mathbf{F}_{1:k}$, as $f_{k|k}(\bm{\mathcal{U}}_k|\mathbf{F}_{1:k})$. Then, deriving the RFS state $\bm{\mathcal{U}}_k$ can be realized by maximizing \emph{a posteriori}.

Note that addressing the posterior density $f_{k|k}(\bm{\mathcal{U}}_k|\mathbf{F}_{1:k})$ is equivalent to addressing its cardinality and its joint elements' PDF. According to Eq. (13), the posterior density can be re-written as:
\begin{equation}
f_{k|k}(\{\mathbf{v}_k\}|\mathbf{F}_{1:k})=\text{Pr}\{|\bm{\mathcal{U}}_k|=1 |\mathbf{F}_{1:k}\}\cdot p(\mathbf{v}_k|\mathbf{F}_{1:k}),
\end{equation}
\begin{equation}
f_{k|k}(\{\mathbf{u}_k,\mathbf{v}_k\}|\mathbf{F}_{1:k}) =\text{Pr}\{|\bm{\mathcal{U}}_k|=2 |\mathbf{F}_{1:k}\}\cdot p(\mathbf{u}_k,\mathbf{v}_k|\mathbf{F}_{1:k}).
\end{equation}

Then, by defining the posterior active probability as:
\begin{equation}
q_{k|k}\triangleq\text{Pr}\{|\bm{\mathcal{U}}_k|=2|\mathbf{F}_{1:k}\},
\end{equation}
and the posterior PDFs as
\begin{equation}
p_{k|k}(\mathbf{v}_k)\triangleq p(\mathbf{v}_k|\mathbf{F}_{1:k}),
\end{equation}
\begin{equation}
p_{k|k}(\mathbf{u}_k,\mathbf{v}_k)\triangleq p(\mathbf{u}_k,\mathbf{v}_k|\mathbf{F}_{1:k}),
\end{equation}
we can estimate the RFS state $\bm{\mathcal{U}}_k$ via pursuing $q_{k|k}$, $p_{k|k}(\mathbf{v}_k)$, and $p_{k|k}(\mathbf{u}_k,\mathbf{v}_k)$.

\subsection{RFS based Two-stage Bayesian Process}
Relying on the transitional density of the RFS state, $\bm{\mathcal{U}}_k$ in Eqs. (16)-(17), and the fluorescence observation trajectory $\mathbf{F}_{1:k}$, two-stage sequential Bayesian process is applied to derive the posterior FISST PDF of $\bm{\mathcal{U}}_k$, i.e.,
\begin{equation}
\begin{aligned}
f_{k|k-1}(\bm{\mathcal{U}}_k|\mathbf{F}_{1:k-1})=&\int\phi_{k|k-1} (\bm{\mathcal{U}}_k|\bm{\mathcal{U}}_{k-1})\\ &\cdot f_{k-1|k-1}(\bm{\mathcal{U}}_{k-1}|\mathbf{F}_{1:k-1})\delta\bm{\mathcal{U}}_{k-1},
\end{aligned}
\end{equation}
\begin{equation}
f_{k|k}(\bm{\mathcal{U}}_k|\mathbf{F}_{1:k})=\frac{\varphi_k(F_k|\bm{\mathcal{U}}_k) \cdot f_{k|k-1}(\bm{\mathcal{U}}_k|\mathbf{F}_{1:k-1})} {\mathlarger{\int}\varphi_k(F_k|\bm{\mathcal{U}}_k) \cdot f_{k|k-1}(\bm{\mathcal{U}}_k|\mathbf{F}_{1:k-1})\delta\bm{\mathcal{U}}_k},
\end{equation}
where Eq. (28) is referred as the \emph{predict-stage}; whilst Eq. (29) accounts for the \emph{update-stage}. Here, $\delta$ represents the set integral rather than the common density integral \cite{6497685, 7423817}.

\subsubsection{Predict-stage}
The predict-stage given by Eq. (28) has an intuitive interpretation. The predicted FISST PDF, $f_{k|k-1}(\bm{\mathcal{U}}_k|\mathbf{F}_{1:k-1})$ generates by evolving the last \emph{posteriori}, $f_{k-1|k-1}(\bm{\mathcal{U}}_{k-1}|\mathbf{F}_{1:k-1})$ through the transitional density $\phi_{k|k-1}(\bm{\mathcal{U}}_k|\bm{\mathcal{U}}_{k-1})$. In this way, the predicted active probability, i.e., $q_{k|k-1}$, and the predicted spatial densities, i.e., $p_{k|k-1}(\mathbf{v}_k)$, and $p_{k|k-1}(\mathbf{u}_k,\mathbf{v}_k)$ are easily computed by introducing them into Eq (28).

\begin{Theo}
The predicted active probability, and the predicted spatial densities can be propagated via:
\begin{equation}
q_{k|k-1}=P_b\cdot(1-q_{k-1|k-1})+P_s(\mathbf{u}_k)\cdot q_{k-1|k-1},
\end{equation}
\begin{equation}
\begin{aligned}
p_{k|k-1}(\mathbf{v}_k)=&\int\pi_{k|k-1}(\mathbf{v}_k|\mathbf{v}_{k-1})\\ &\cdot p_{k-1|k-1}(\mathbf{v}_{k-1})d\mathbf{v}_{k-1},
\end{aligned}
\end{equation}
\begin{equation}
\begin{aligned}
p_{k|k-1}(\mathbf{u}_k,&\mathbf{v}_k)=\frac{P_b\cdot(1-q_{k-1|k-1})}{q_{k|k-1}}\cdot b_k(\mathbf{u}_k)\cdot p_{k|k-1}(\mathbf{v}_k)\\
&+\frac{P_s(\mathbf{u}_k)\cdot q_{k-1|k-1}}{q_{k|k-1}}\tiny{\iint} p_{k-1|k-1}(\mathbf{u}_{k-1},\mathbf{v}_{k-1})\\
&\cdot \psi_{k|k-1}(\mathbf{u}_k,\mathbf{v}_k |\mathbf{u}_{k-1},\mathbf{v}_{k-1}) d\mathbf{u}_{k-1}\mathbf{v}_{k-1}.
\end{aligned}
\end{equation}
\end{Theo}

From \emph{Theorem 1} (see detail in \emph{Appendix A}), three aspects are considered in propagating $q_{k|k-1}$, $p_{k|k-1}(\mathbf{v}_k)$ and $p_{k|k-1}(\mathbf{u}_k,\mathbf{v}_k)$.

i) For the predicted active probability $q_{k|k-1}$, it should contain two complementary terms, i.e., the active-to-active and the silent-to-active signal sequence.

ii) For the predicted density $p_{k|k-1}(\mathbf{v}_k)$, it is the transitional effect that evolves the posterior density from the previous $p_{k-1|k-1}(\mathbf{v}_{k-1})$, since $\mathbf{v}_k$ cannot disappear from the fluorescence observations, i.e., whether the spike is absent or present at the time $k$, we need to estimate $v_k$ all the time.

iii) For the predicted density $p_{k|k-1}(\mathbf{u}_k,\mathbf{v}_k)$, we should consider the evolutions from the silent-to-active pieces, representing by the new activated density, as well as the active-to-active term.

\begin{Prop}
The predicted spatial PDF $p_{k|k-1}(\mathbf{u}_k,\mathbf{v}_k)$ can be propagated separately as:
\begin{equation}
p_{k|k-1}(\mathbf{u}_k,\mathbf{v}_k)=p_{k|k-1}(\mathbf{u}_k)\cdot p_{k|k-1}(\mathbf{v}_k),
\end{equation}
where
\begin{equation}
\begin{aligned}
p_{k|k-1}(\mathbf{u}_k)=&\frac{P_b\cdot(1-q_{k-1|k-1})}{q_{k|k-1}}\cdot b_k(\mathbf{u}_k)\\
&+\frac{P_s(\mathbf{u}_k)\cdot q_{k-1|k-1}}{q_{k|k-1}}\tiny{\int} p_{k-1|k-1}(\mathbf{u}_{k-1})\\
&\cdot \varpi_{k|k-1}(\mathbf{u}_k|\mathbf{u}_{k-1})d\mathbf{u}_{k-1}.
\end{aligned}
\end{equation}
\end{Prop}

From \emph{Proposition 1}, we can respectively compute the predicted PDFs, i.e., $p_{k|k-1}(\mathbf{u}_k)$, and $p_{k|k-1}(\mathbf{v}_k)$. In this way, we are allowed to reduce the parameter space of dimension 6 to two parameter spaces of dimension 3, thereby potentially reducing the computation complexity.

\subsubsection{Update-stage}
In the update-stage given by Eq. (29), the posterior FISST PDF $f_{k|k}(\bm{\mathcal{U}}_k|\mathbf{F}_{1:k})$ is computed by updating the predicted density, $f_{k|k-1}(\bm{\mathcal{U}}_k|\mathbf{F}_{1:k-1})$, via the current likelihood PDF, $\varphi_k(F_k|\bm{\mathcal{U}}_k)$. Accordingly, the posterior active probability, i.e., $q_{k|k}$, and the posterior spatial densities, i.e., $p_{k|k}(\mathbf{v}_k)$, and $p_{k|k}(\mathbf{u}_k,\mathbf{v}_k)$ can be respectively derived with the following \emph{Theorem}.

\begin{Theo}
The posterior active probability, and the spatial densities are updated by:
\begin{equation}
\begin{aligned}
&q_{k|k}\\
=&q_{k|k-1}\iint\varphi_k(F_k|\{\mathbf{u}_k,\mathbf{v}_k\})\cdot p_{k|k-1}(\mathbf{u}_k,\mathbf{v}_k)d\mathbf{u}_k\mathbf{v}_k\\
&\cdot\bigg(q_{k|k-1}\iint\varphi_k(F_k|\{\mathbf{u}_k,\mathbf{v}_k\})\cdot p_{k|k-1}(\mathbf{u}_k,\mathbf{v}_k)d\mathbf{u}_k\mathbf{v}_k\\
&+(1-q_{k|k-1})\int\varphi_k(F_k|\{\mathbf{v}_k\})\cdot p_{k|k-1}(\mathbf{v}_k)d\mathbf{v}_k\bigg)^{-1},
\end{aligned}
\end{equation}
and,
\begin{equation}
p_{k|k}(\mathbf{u}_k,\mathbf{v}_k)=\frac{\varphi_k(F_k|\{\mathbf{u}_k,\mathbf{v}_k\})\cdot p_{k|k-1}(\mathbf{u}_k,\mathbf{v}_k)}{\iint\varphi_k(F_k|\{\mathbf{u}_k,\mathbf{v}_k\}) p_{k|k-1}(\mathbf{u}_k,\mathbf{v}_k)d\mathbf{u}_k\mathbf{v}_k},
\end{equation}
\begin{equation}
p_{k|k}(\mathbf{v}_k)=\int p_{k|k}(\mathbf{u}_k,\mathbf{v}_k)d\mathbf{u}_k.
\end{equation}
\end{Theo}

An intuitive description of Eq. (35) is the proportion of active density in the whole active and silent signal sequence. Eq. (36) is straightforward by updating the predicted densities through the likelihood. Eq. (37) is the marginal distribution of (36). Note that $p_{k|k}(\mathbf{v}_k)$ in Eq. (37) is also suitable for the case of silent signal, and will be degenerated to the traditional Bayesian form as:
\begin{equation}
p_{k|k}(\mathbf{v}_k)=\frac{\varphi_k(F_k|\{\mathbf{v}_k\})\cdot p_{k|k-1}(\mathbf{v}_k)}{\int\varphi_k(F_k|\{\mathbf{v}_k\})\cdot p_{k|k-1}(\mathbf{v}_k)d\mathbf{v}_k},
\end{equation}
if $\bm{\mathcal{U}}_k=\{\mathbf{v}_k\}$. The proof of \emph{Theorem 2} is specified in \emph{Appendix B}.

Given the computations of the posterior densities, the optimal estimations of $\mathbf{u}_k$, and $\mathbf{v}_k$ can be realized via maximizing \emph{a posteriori}, i.e.,
\begin{equation}
\hat{\mathbf{u}}_k=
\begin{cases}
\begin{aligned}
\underset{\mathbf{u}_k}{\argmax}\int p_{k|k}(\mathbf{u}_k,\mathbf{v}_k)d\mathbf{v}_k&\\
& q_{k|k}\geq q_\text{th},\\
\bigg[\exp(-\frac{\Delta_t}{\hat{\tau}_{k-1|k-1}})\hat{C}_{k-1|k-1}, &\hat{\tau}_{k-1|k-1},\hat{A}_{k-1|k-1}\bigg]^T\\
& q_{k|k}<q_\text{th},
\end{aligned}
\end{cases}
\end{equation}
\begin{equation}
\hat{\mathbf{v}}_k=\underset{\mathbf{v}_k}{\argmax}~p_{k|k}(\mathbf{v}_k),
\end{equation}
where $q_\text{th}$ is a pre-defined threshold empirically
configured to $q_\text{th}=0.5$.

Relying on Eqs. (39)-(40), the spike detection is equivalent to examining the difference of $\hat{C}_{k|k}$ and $\hat{C}_{k-1|k-1}$. Baseline is tracked with the acquisition of $\hat{B}_{k|k}$. Model parameters i.e., $\hat{A}$, $\hat{\tau}$, $\hat{\sigma}$ and $\hat{\eta}$ can be derived through $\hat{\mathbf{u}}_k$ and $\hat{\mathbf{v}}_k$.

\subsection{Implementation of RFS-Bayesian Scheme}
Note from Eqs. (31)-(32) that the derivations of predicted spatial
densities $p_{k|k-1}(\mathbf{v}_k)$, and $p_{k|k-1}(\mathbf{u}_k,\mathbf{v}_k)$ rely on the marginalization of continuous distributions. As far as the integration on the high dimensional density is concerned (e.g. 6-dimension in the case of $|\bm{\mathcal{U}}_k|=2$), the implementation may be computationally intractable. To cope with this difficulty, a numerical approach, i.e., particle filter (PF), is suggested to alleviate the computational complexity of the RFS-Bayesian inference.

For short, PF is a simulated numerical method, which approximates the complex distribution, (e.g., $p(\mathbf{u})$) via a group of random discrete particles, denote as $\mathbf{u}^{(i)}$ with their probability weights, denoted as $w^{(i)}(i=1,2,...,I)$, i.e., $p(\mathbf{u})\backsimeq\sum_{i=0}^I\delta(\mathbf{u}-\mathbf{u}^{(i)})\times w^{(i)}$.

\subsubsection{RFS-Bayesian PF}
For the implementation of the RFS-Bayesian scheme, the PF is used to approximate the predicted spatial densities $p_{k|k-1}(\mathbf{v}_k)$, and $p_{k|k-1}(\mathbf{u}_k,\mathbf{v}_k)$, i.e,
\begin{equation}
p_{k|k-1}(\mathbf{v}_k)\simeq\sum_{i=1}^Iw_{k|k-1}^{(i)} \delta(\mathbf{v}_k-\mathbf{v}_{k|k-1}^{(i)}),
\end{equation}
\begin{equation}
p_{k|k-1}(\mathbf{u}_k,\mathbf{v}_k)\simeq\sum_{i=1}^Iw_{k|k-1}^{(i)} \delta(\mathbf{u}_k-\mathbf{u}_{k|k-1}^{(i)})\delta(\mathbf{v}_k-\mathbf{v}_{k|k-1}^{(i)}),
\end{equation}
where $\mathbf{u}_{k|k-1}^{(i)}(i=1,2,...,I)$ are the $i$th particles denote all possible $\mathbf{u}_k$; whilst $\mathbf{v}_{k|k-1}^{(i)}$ denote all possible $\mathbf{v}_k$. $w_{k|k-1}^{(i)}$ is the weight for particle $[\mathbf{u}_{k|k-1}^{(i)},\mathbf{v}_{k|k-1}^{(i)}]^T$. In this view, the purpose of approximating $p_{k|k-1}(\mathbf{v}_k)$, and $p_{k|k-1}(\mathbf{u}_k,\mathbf{v}_k)$ in Eq. (41)-(42) is equivalent to deriving the predicted particles and their weights, i.e., $[\mathbf{u}_{k|k-1}^{(i)},\mathbf{v}_{k|k-1}^{(i)}]^T$, and $w_{k|k-1}^{(i)}$.

Recalling from Eqs. (31)-(32), $p_{k|k-1}(\mathbf{v}_k)$ comprises only the survival component; whilst $p_{k|k-1}(\mathbf{u}_k,\mathbf{v}_k)$ is composed of both a birth component and the survival component. Hence, three proposal distributions are designed to simulate respectively the two particles $\mathbf{u}_{k|k-1}^{(i)}$, and $\mathbf{v}_{k|k-1}^{(i)}$, i.e.,
\begin{eqnarray}
\mathbf{u}_{k|k-1}^{(i)}&\sim&
\begin{cases}
\begin{aligned}
\beta_k(\mathbf{u}_k|\mathbf{F}_{1:k-1}),\\
& i=1,2,...,B\\
\xi_k(\mathbf{u}_k|\mathbf{u}_{k-1|k-1}^{(i)}&,\mathbf{F}_{1:k-1}),\\
& i=B+1,B+2,...,I\\
\end{aligned}
\end{cases}
\\[3mm]
\mathbf{v}_{k|k-1}^{(i)}&\sim&\varrho_k(\mathbf{v}_k |\mathbf{v}_{k-1|k-1}^{(i)},\mathbf{F}_{1:k-1})
\end{eqnarray}
In Eq. (43), the first $B$ particles are used to approximate the activated term, i.e., $b_k(\mathbf{u}_k)$ in the silent-to-active case in Eq. (32); whilst the rest particles simulate the other active-to-active component. In Eq. (44), total $I$ particles are applied in depicting the predicted PDF in Eq. (31).

Given total $I$ simulated particles $[\mathbf{u}_{k|k-1}^{(i)},\mathbf{v}_{k|k-1}^{(i)}]^T$, the weights thereby can be computed as:
\begin{equation}
\begin{aligned}
&w_{k|k-1}^{(i)}\\
=&
\begin{cases}
\begin{aligned}
p_{b,k}\cdot&\frac{b_k(\mathbf{u}_{k|k-1}^{(i)})\pi_{k|k-1} (\mathbf{v}_{k|k-1}^{(i)}|\mathbf{v}_{k-1|k-1}^{(i)})} {\beta_k(\mathbf{u}_{k|k-1}^{(i)}|\mathbf{F}_{1:k-1}) \varrho_k(\mathbf{v}_{k|k-1}^{(i)}|\mathbf{v}_{k-1|k-1}^{(i)},\mathbf{F}_{1:k-1})}\\[2mm]
& i=1,2,...,B\\
p_{s,k}^{(i)}\cdot&\frac{\psi_{k|k-1} (\mathbf{u}_{k|k-1}^{(i)},\mathbf{v}_{k|k-1}^{(i)}|\mathbf{u}_{k-1|k-1}^{(i)},\mathbf{v}_{k-1|k-1}^{(i)})} {{(\xi\varrho)}_k (\mathbf{u}_{k|k-1}^{(i)},\mathbf{v}_{k|k-1}^{(i)} |\mathbf{u}_{k-1|k-1}^{(i)},\mathbf{v}_{k-1|k-1}^{(i)},\mathbf{F}_{1:k-1})}\\[2mm]
& i=B+1,B+2,...,I
\end{aligned}
\end{cases}
\end{aligned}
\end{equation}
where the $p_{b,k}=P_b\cdot(1-q_{k-1|k-1})/q_{k|k-1}/B$, and the $p_{s,k}^{(i)}=w_{k-1|k-1}^{(i)}\cdot P_s(\mathbf{u}_{k|k-1}^{(i)})\cdot q_{k-1|k-1}/q_{k|k-1}$. Sign $(\xi\varrho)_k(\cdot)$ denotes the multiplication of $\xi_k(\cdot)$ and $\varrho_k(\cdot)$.

Once the proposal densities, i.e., $\beta_k(\mathbf{u}_k|\mathbf{F}_{1:k-1})$, $\xi_k(\mathbf{u}_k|\mathbf{u}_{k-1|k-1}^{(i)},\mathbf{F}_{1:k-1})$ and $\varrho_k(\mathbf{v}_k |\mathbf{v}_{k-1|k-1}^{(i)},\mathbf{F}_{1:k-1})$ are specified properly (in the next two parts), the predict-stage of the two-stage Bayesian process is realized, and we derive the predicted particles and weights, i.e., $\big\{[\mathbf{u}_{k|k-1}^{(i)},\mathbf{v}_{k|k-1}^{(i)}]^T,w_{k|k-1}^{(i)}\big\}_{i=1}^I$.
Then, the update-stage given by Eqs. (36)-(37) can be pursued by updating the weights through likelihoods, i.e.,
\begin{equation}
w_{k|k}^{(i)}=\frac{\varphi_k(F_k|\{\mathbf{u}_{k|k}^{(i)},\mathbf{v}_{k|k}^{(i)}\}) \cdot w_{k|k-1}^{(i)}}{\sum_{i=1}^I\varphi_k(F_k|\{\mathbf{u}_{k|k}^{(i)},\mathbf{v}_{k|k}^{(i)}\}) \cdot w_{k|k-1}^{(i)}},
\end{equation}
with $\mathbf{u}_{k|k}^{(i)}=\mathbf{u}_{k|k-1}^{(i)}$, and $\mathbf{v}_{k|k}^{(i)}=\mathbf{v}_{k|k-1}^{(i)}$.

With the help of Eq. (46) in deriving the updated particles and their weights, we can simulate both the posterior active probability (i.e., $q_{k|k}$), and the posterior density (i.e., $p_{k|k}(\mathbf{v}_k)$, and $p_{k|k}(\mathbf{u}_k,\mathbf{v}_k)$). The posterior active probability is computed as:
\begin{equation}
q_{k|k}=\frac{q_{k|k-1}\sum_{i=1}^Iw_{k|k}^{(i)}}{q_{k|k-1}\sum_{i=1}^Iw_{k|k}^{(i)} +(1-q_{k|k-1})\sum_{i=1}^I\tilde{w}_{k|k}^{(i)}}
\end{equation}
where
\begin{equation}
\tilde{w}_{k|k}^{(i)}=\frac{\varphi_k(F_k|\{\mathbf{v}_{k|k}^{(i)}\}) \cdot w_{k|k-1}^{(i)}}{\sum_{i=1}^I\varphi_k(F_k|\{\mathbf{v}_{k|k}^{(i)}\}) \cdot w_{k|k-1}^{(i)}}.
\end{equation}
The spatial densities (i.e., $p_{k|k}(\mathbf{v}_k)$, and $p_{k|k}(\mathbf{u}_k,\mathbf{v}_k)$), are approximated by:
\begin{equation}
p_{k|k}(\mathbf{v}_k)\simeq\sum_{i=1}^Iw_{k|k}^{(i)} \delta(\mathbf{v}_k-\mathbf{v}_{k|k}^{(i)}),
\end{equation}
\begin{equation}
p_{k|k}(\mathbf{u}_k,\mathbf{v}_k)\simeq\sum_{i=1}^Iw_{k|k}^{(i)} \delta(\mathbf{u}_k-\mathbf{u}_{k|k}^{(i)})\delta(\mathbf{v}_k-\mathbf{v}_{k|k}^{(i)}),
\end{equation}

At the end of each iteration, a re-sample procedure should be adopted in order to eliminate particles with negligible weights \cite{1236770, 7423817, 6497685}.

\subsubsection{Proposal Active-to-active Densities}
The design of the active-to-active densities in Eqs. (43)-(44), i.e., $\xi_k(\mathbf{u}_k|\mathbf{u}_{k-1|k-1}^{(i)},\mathbf{F}_{1:k-1})$, and $\varrho_k(\mathbf{v}_k |\mathbf{v}_{k-1|k-1}^{(i)},\mathbf{F}_{1:k-1})$ is straightforward. As we know the transitional PDFs as $\varpi_{k|k-1}(\mathbf{u}_k|\mathbf{u}_{k-1})$, and $\pi_{k|k-1}(\mathbf{v}_k|\mathbf{v}_{k-1})$, we can directly assign these transitional PDFs as the proposals of the active-to-active densities, i.e.,
\begin{equation}
\xi_k(\mathbf{u}_k|\mathbf{u}_{k-1|k-1}^{(i)},\mathbf{F}_{1:k-1}) \triangleq\varpi_{k|k-1}(\mathbf{u}_k|\mathbf{u}_{k-1|k-1}^{(i)}),
\end{equation}
\begin{equation}
\varrho_k(\mathbf{v}_k |\mathbf{v}_{k-1|k-1}^{(i)},\mathbf{F}_{1:k-1}) \triangleq\pi_{k|k-1}(\mathbf{v}_k|\mathbf{v}_{k-1|k-1}^{(i)}).
\end{equation}
One may refer to \cite{1236770, 7423817} for details.

\subsubsection{Proposal Silent-to-active Density}
The proposal of the silent-to-active density, i.e., $\beta_k(\mathbf{u}_k|\mathbf{F}_{1:k-1})$ heavily relies on the design of the activated density, i.e., $b_k(\mathbf{u}_k)$. To simplify it, we here define the proposal $\beta_k(\mathbf{u}_k|\mathbf{F}_{1:k-1})$ as follows:
\begin{equation}
\beta_k(\mathbf{u}_k|\mathbf{F}_{1:k-1})\triangleq b_k(\mathbf{u}_k).
\end{equation}

For the activated density, i.e., $b_k(\mathbf{u}_k)$, the major challenge is how to effectively address the likelihood disappearance, aroused by the silent piece (i.e., $q_{k|k}\rightarrow0$). If this happens, there is no likelihood information can be used to infer $\mathbf{u}_k$. More specifically, given the threshold as $q_\text{th}$, consider that the active probability at the time-slot $k-J$ is the lasted one that is no lesser than $q_\text{th}$, i.e., $q_{k-J|k-J}\geq q_\text{th}~\text{and}~ q_{k-j|k-j}<q_\text{th}, 0\leq j<J$. In this case, the likelihood sequence i.e., $\varphi_{k-J+1:k}(\cdot)$ contains no information when estimating $\mathbf{u}_{k-J+1:k}$, and therefore the inference of \emph{a posteriori} density will be infeasible. In order to cope with this difficult problem, we design an adaptive birth density, premised on the lasted state whose active probability is no lesser than $q_\text{th}$ i.e.,
\begin{equation}
\begin{aligned}
b_k(\mathbf{u}_k)=\underbrace{\iint\cdots\int}_{J}& b_{k-J}(\mathbf{u}_{k-J})\cdot\\ &\prod_{j=1}^J\varpi_{k-j+1|k-j}(\mathbf{u}_{k-j+1}|\mathbf{u}_{k-j})d\mathbf{u}_{k-j}.
\end{aligned}
\end{equation}

\subsubsection{Algorithm Flow}
\begin{figure*}[!t]
\centering
\includegraphics[width=6in]{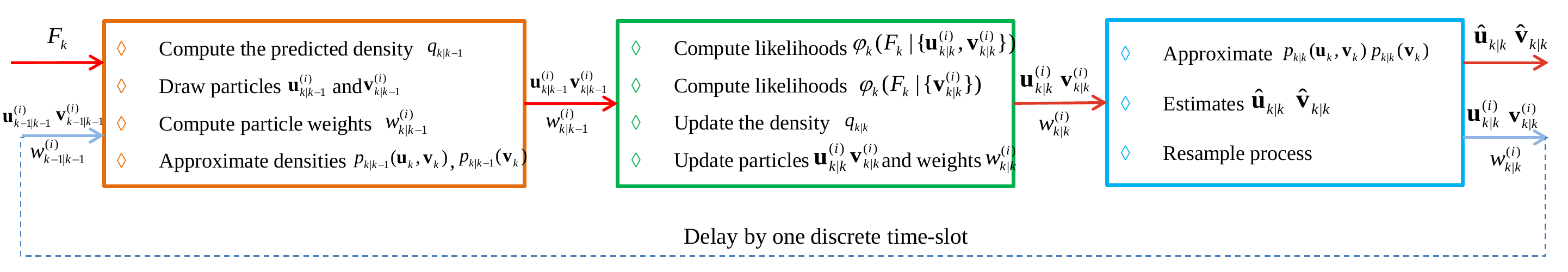}
\caption{Schematic algorithm flow of the PF-based RFS Bayesian scheme.}
\end{figure*}
A schematic flow of the proposed RFS-Bayesian scheme is illustrated by Fig. 3. At the time-slot $k$, the particles $\mathbf{u}_{k-1|k-1}^{(i)},\mathbf{v}_{k-1|k-1}^{(i)}$
and associated weights $w_{k-1|k-1}^{(i)}$ serve as the inputs, as well as the current fluorescence observation $F_k$. Steps 1-3 aim to
derive the predicted particles and their weights. Step 4 is to approximate the predicted densities, $p_{k|k-1}(\mathbf{v}_k)$, and $p_{k|k-1}(\mathbf{u}_k,\mathbf{v}_k)$. Then, steps 5-6 compute the likelihoods, and update the particles and their weights. Steps 7-8 calculate the posterior \emph{active} probability, $q_{k|k}$, and the posterior densities, $p_{k|k}(\mathbf{v}_k)$, and $p_{k|k}(\mathbf{u}_k,\mathbf{v}_k)$. Step 9 estimates the $\hat{\mathbf{u}}_k$, and $\hat{\mathbf{v}}_k$. Step 10 updates $J$ for generate birth density. Finally, step 11 implements the re-sample process, which not only preserves the diversity of particles, but also eliminates useless particles with negligible weights. Step 12 normalizes the weights of re-sampled particles. The outputs contain the estimation of $\hat{\mathbf{u}}_k$ and $\hat{\mathbf{v}}_k$,, as well as the particles at time-slot $k$, which will be utilized in the next-step (i.e. time-slot $k+1$) estimation.

\section{Numerical Simulations}
\begin{figure*}[!t]
\centering
\subfloat[Estimation of $A$.]{\includegraphics[width=2.5in]{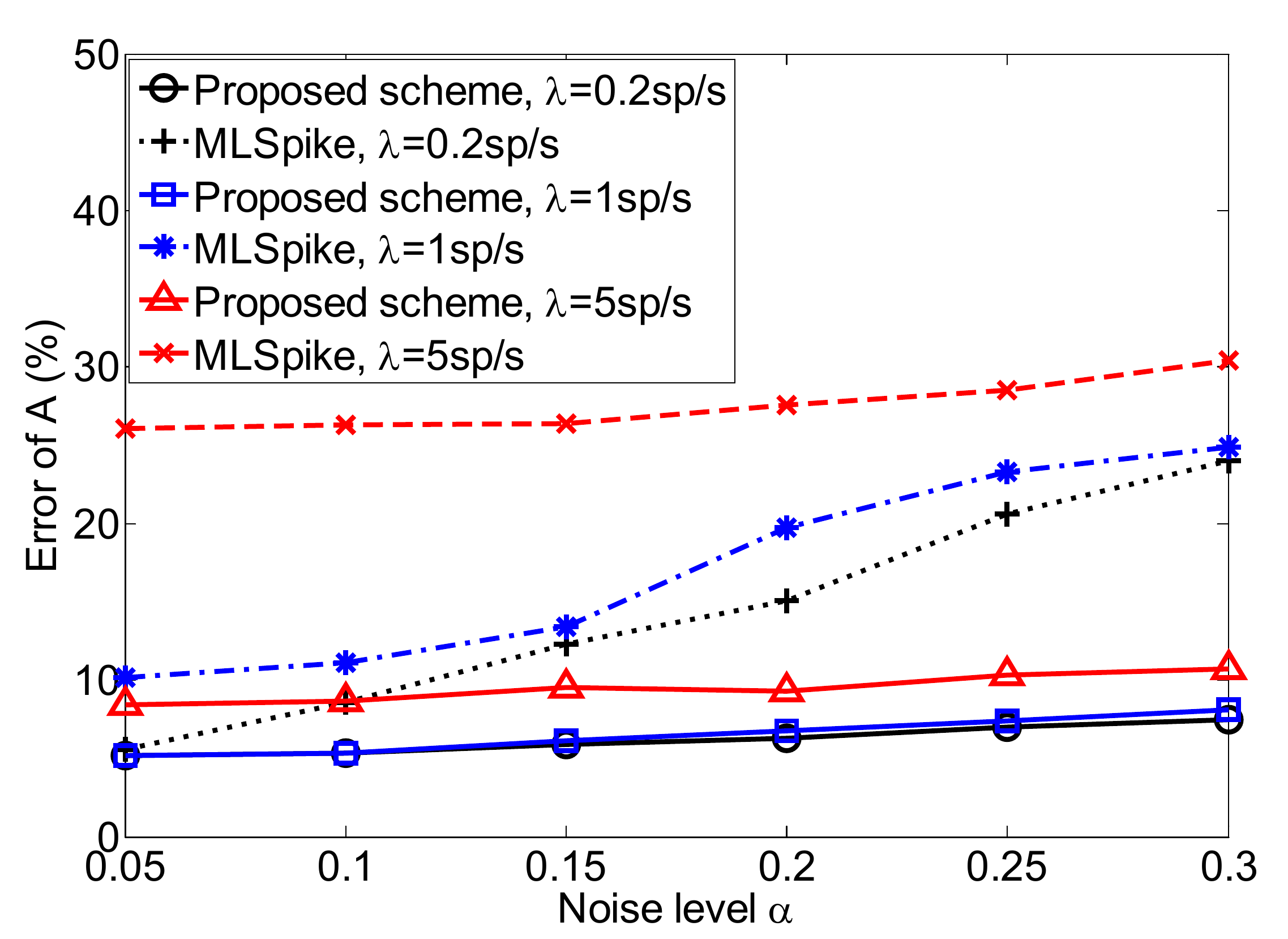}}
\hfil
\subfloat[Estimation of $\tau$.]{\includegraphics[width=2.5in]{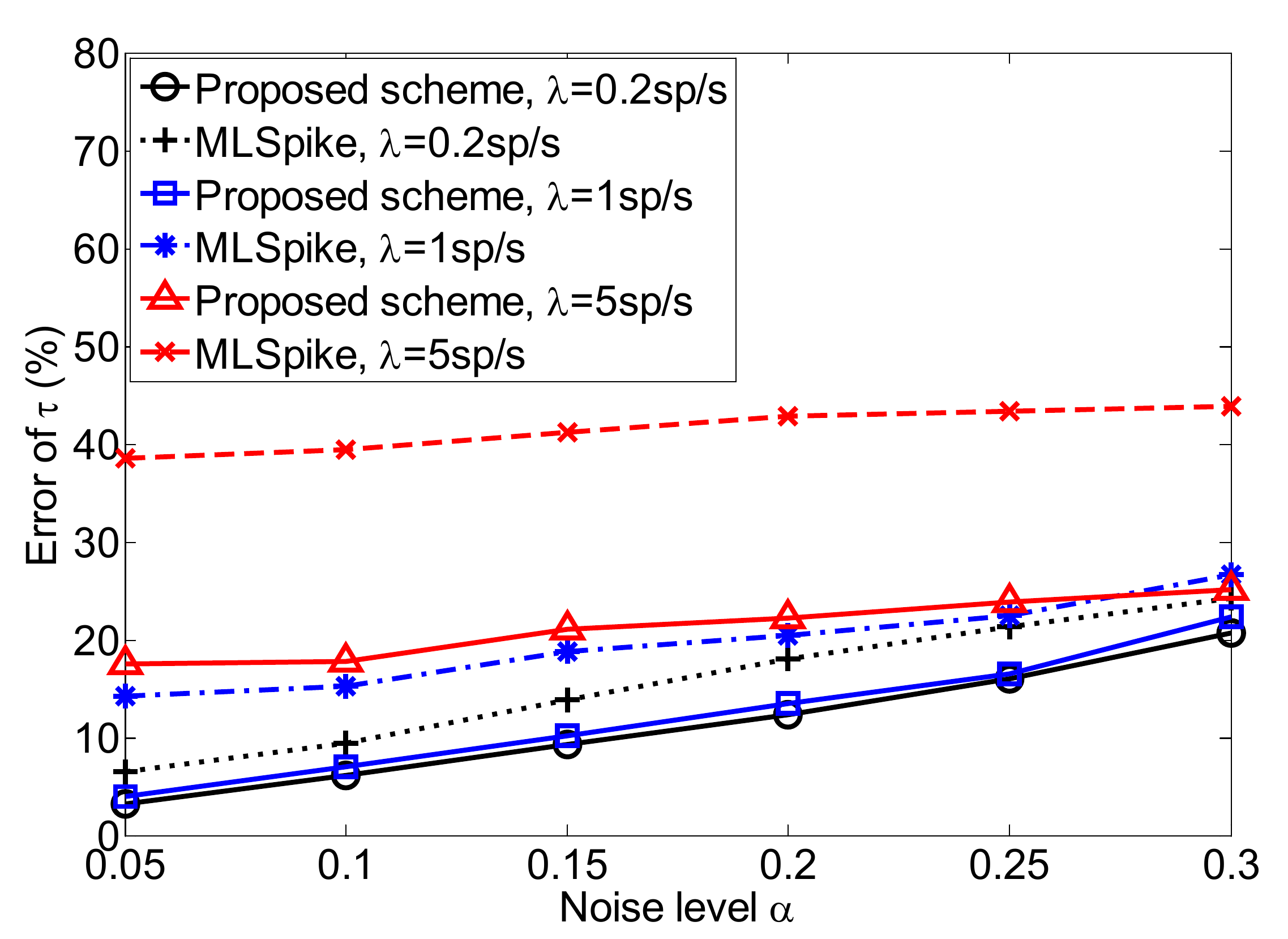}}
\caption{Performance on estimating model parameters}
\end{figure*}

In the following analysis, the performance of our RFS-Bayesian scheme will be evaluated, in terms of the accuracy of spike/baseline inferences and estimations of model parameters, as well as the computational complexity, in contrast with the state-of-the-art MLSpikes \cite{Deneux2016Accurate}. First, the estimation performance of the spike-related parameters (i.e., amplitude $A$, and time-delay $\tau$) is illustrated, as well as the MLspike algorithm integrated the auto-calibration mechanisms. Then, the accuracy of estimations of background-related parameters (i.e., the variances of ambient noise, i.e., $\sigma^2$, and process noise of baseline, i.e., $\eta^2$) is evaluated. Finally, we evaluate the detection/estimation performance of both spikes and baseline.

The involved parameters in numerical simulations are configured as follows. As far as the neuron signalling scenario is considered, various spike rate, i.e., $\lambda=0.2/$s, $\lambda=1/$s, and $\lambda=5/$s are evaluated, given the sample interval as $\Delta_t=0.02$s. Without the loss of generality, we randomly assign the time-delay constant as $\tau\in[\tau_{\text{min}},\tau_{\text{max}}]=[0.6,1]$ \cite{Deneux2016Accurate}. For the fluorescence measurement function, the reference fluorescence at rest is $F_0=1$. The transient amplitude $A$ is assigned randomly within the interval $[A_{\text{min}},A_{\text{max}}]=[0.04,0.1]$. The saturation parameter is $\gamma=0.1$ \cite{Deneux2016Accurate}. The ambient noise, normalized by the transient amplitude $A$, is configured as $\sigma=\alpha\cdot A$, where $\alpha$ ranges from $0.01$ to $0.3$.

\subsection{Estimation Performance of Spike-related Parameters}
The evaluation of RFS-Bayesian scheme on estimating spike-related model parameters, i.e., $A$ and $\tau$, is illustrated in Fig. 4, measured via the relative error, i.e.,
\begin{equation}
\text{RE}_A\triangleq\mathbb{E}\bigg\{\frac{||\hat{A}-A||_2}{A}\bigg\},
\end{equation}
\begin{equation}
\text{RE}_\tau\triangleq\mathbb{E}\bigg\{\frac{||\hat{\tau}-\tau||_2}{\tau}\bigg\}.
\end{equation}

\subsubsection{Estimation Performance of Transient Amplitude}
The accuracy of estimating transient amplitude (i.e., $A$) is shown in Fig. 4-(a). we can observe that as the ambient noise level (i.e., $\alpha=\sigma/A$) rises, the estimating error $\text{RE}_A$ increases, demonstrating the performance deterioration caused by the ambient noise. Another observation is that with the increment of spike rate (i.e., $\lambda$), it becomes more difficult to accurately estimate $A$. For instance, given a fixed noise level $\alpha=0.2$, the $\text{RE}_A$ increases from $10\%$ to $27\%$ as the spike rate $\lambda$ grows from $0.2$ to $5$. This is mainly because a higher spike rate introduces more interactions between the spikes and baseline, which makes the accurate estimation/detection difficult.

By further comparing the estimating errors of the proposed RFS-Bayesian scheme and the state-of-the-art MLSpike, we see that the $\text{RE}_A$ from the proposed scheme keeps lower than that of the MLSpike, suggesting that the proposed scheme effectively improves over the MLSpike, and is more robust to the noise and interactions. This is attributed to the capability of the RFS state to distinguish the active/silent pieces, so the proposed scheme can use only the active piece to estimate the spike-related amplitude $A$, and simultaneously refine the parameter as the recursive estimation advances. By contrast, the auto-calibration embedded in the MLSpike may use false spikes in silent pieces to estimate spike-related parameters, therefore leading to erroneous $A$.

\subsubsection{Estimation Performance of Estimating Time-delay}
The estimation performance of the time-delay $\tau$ is shown in Fig. 4-(b). It is also intuitive to see that the $\text{RE}_\tau$ grows with both the noise level $\alpha$, and the spike rate $\lambda$ increasing.

Then, in contrast with the MLSpike scheme, the estimation performance of the proposed RFS-Bayesian scheme is better. For instance, at low spike rates (e.g., $\lambda=0.2$sp/s and $\lambda=1$sp/s), $\text{RE}_\tau$ from the proposed scheme is smaller than that of the MLSpike. This advantage becomes even greater in the case of a higher spike rate (i.e., $\lambda=5$sp/s). The $\text{RE}_\tau$ from the proposed scheme, reaches nearly $22\%$, a half smaller than the value of the MLSpike (about $42\%$). The accuracy benefit of the proposed RFS-Bayesian is mainly due to the capability of distinguishing the active/silent pieces of observations. As such, we avoid using the silent piece to estimate spike-related parameters (i.e., $\tau$ and  $A$), and thereby are able to implement reliable parameter estimation.

\subsection{Estimation Performance of Background-related Parameters}
\begin{figure*}[!t]
\centering
\subfloat[Estimation of $\sigma$.]{\includegraphics[width=2.5in]{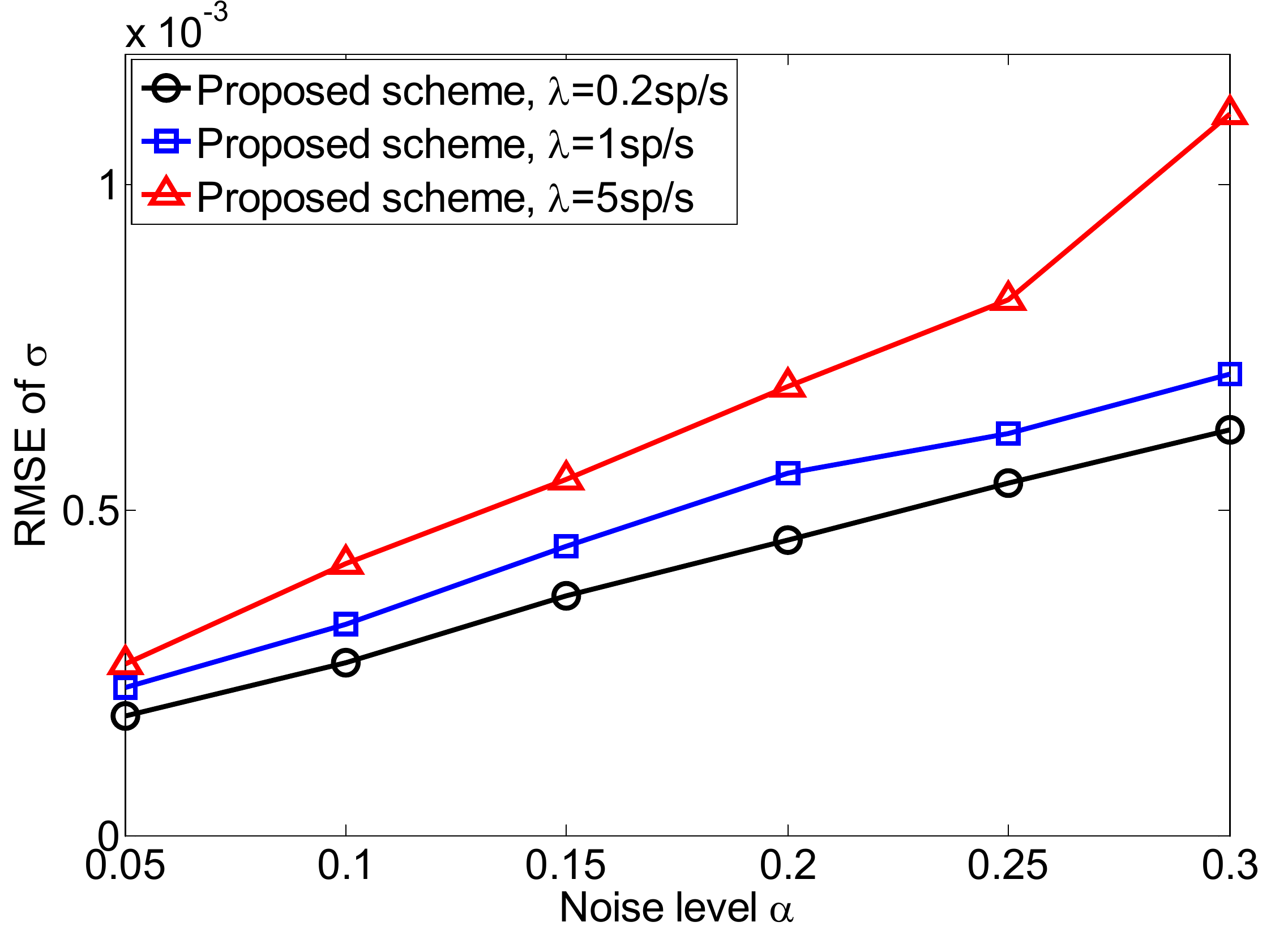}}
\hfil
\subfloat[Estimation of $\eta$.]{\includegraphics[width=2.5in]{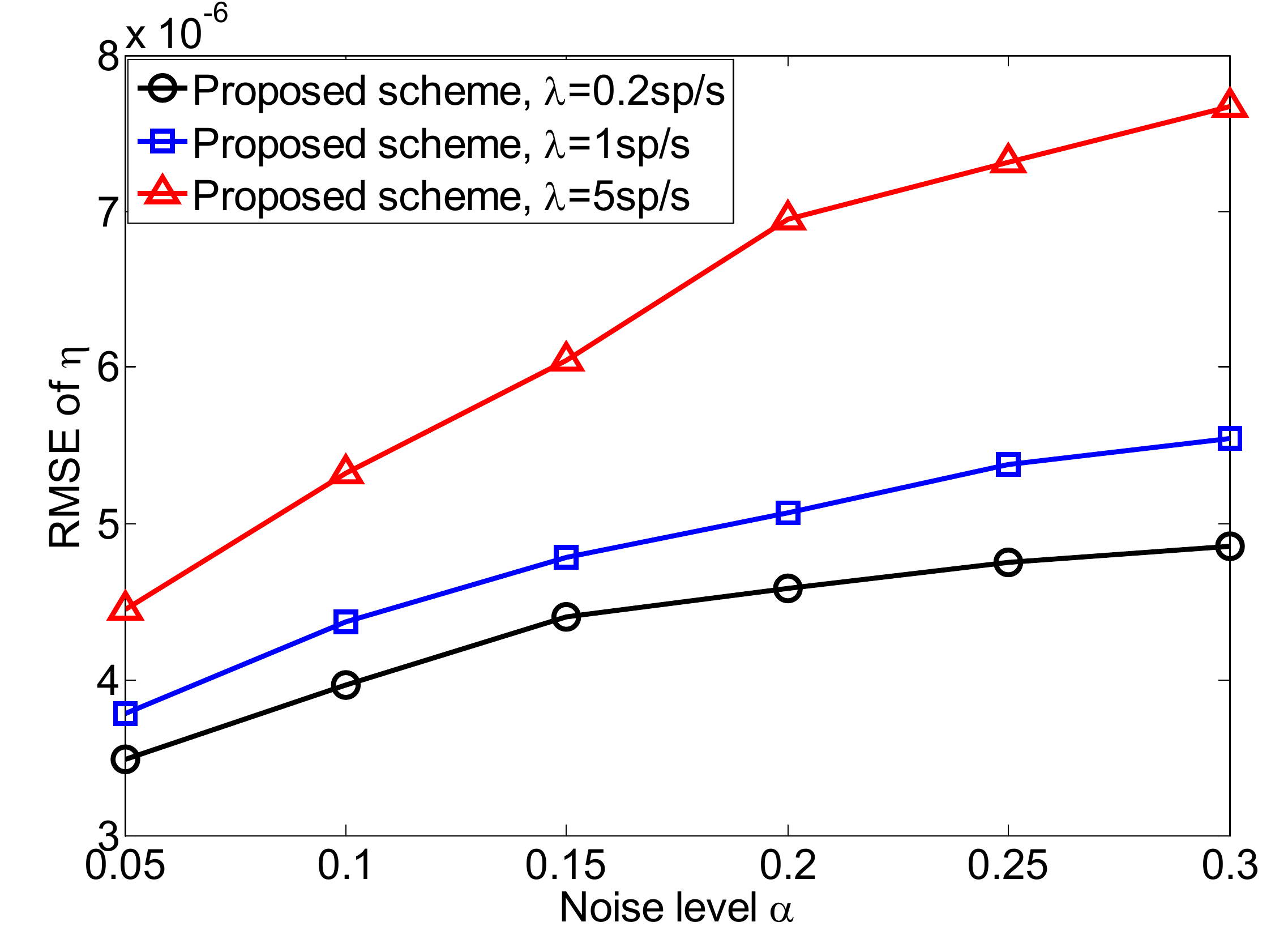}}
\caption{Performance on estimating variances of ambient noise and process noise.}
\end{figure*}

The performance of RFS-Bayesian scheme on estimating background-related parameters (i.e., the variances of ambient noise $\sigma^2$ and baseline $\eta^2$) are shown in Fig. 5. Here, we measure the accuracy in terms of the root mean square error (RMSE) of $\sigma$ and $\eta$, i.e.,
\begin{equation}
\text{RMSE}_\sigma\triangleq\mathbb{E}\{||\hat{\sigma}-\sigma||_2\},
\end{equation}
\begin{equation}
\text{RMSE}_\eta\triangleq\mathbb{E}\{||\hat{\eta}-\eta||_2\}.
\end{equation}

\subsubsection{Estimation Performance of Noise Variance}
From Fig. 5-(a), the $\text{RMSE}_\sigma$ is evaluated in various noise levels and spike rates. It is illustrated that the estimation performance is deteriorated by the growth of the ambient noise $\alpha$ and spike rate $\lambda$. For instance, at a fixed spike rate $\lambda=1$sp/s, $\text{RMSE}_\sigma$ increases from $0.2\times10^{-3}$ to almost $0.7\times10^{-3}$ as $\alpha$ rises. And given a fixed noise level $\alpha=0.3$, $\text{RMSE}_\sigma$ increases from $0.5\times10^{-3}$ to about $1.6\times10^{-3}$ when the spike rate $\lambda$ rises.

Note that although the estimation performance is undermined by the increased ambient noise and spike rate, the value of $\text{RMSE}_\sigma$ is still acceptable. As a matter of fact, the relative error of $\sigma$, i.e., $\text{RE}_\sigma=\mathbb{\text{RMSE}_\sigma/\sigma}$ is no more than $8\%$, which gives a fair estimation of $\sigma$ for practical uses.

\subsubsection{Estimation Performance of Baseline Variance}
The estimation accuracy of baseline variance is shown in Fig. 5-(b). We see that the $\text{RMSE}_\eta$ is also severely subjected to the increments of both the ambient noise and the spike rate. For instance, given a fixed spike rate $\lambda=0.2$sp/s, the $\text{RMSE}_\eta$ rises from $3.5\times10^{-6}$ to about $4.9\times10^{-6}$ as the noise level grows from $0.05$ to $0.3$. When the spike rate increases from $0.2$sp/s to $5$sp/s, the $\text{RMSE}_\eta$ increases from $4\times10^{-6}$ to $5.3\times10^{-6}$ at a fixed noise level $\alpha=0.1$.

For the spike detection application, the performance given by Fig. 5-(b) is still tolerant. Its relative error, computed as $\text{RE}_\eta=\mathbb{\text{RMSE}_\eta/\eta}$, is almost lesser than $0.55\%$, demonstrating a reliable estimation result for subsequently usages.

\begin{figure*}[!t]
\centering
\includegraphics[width=5.5in]{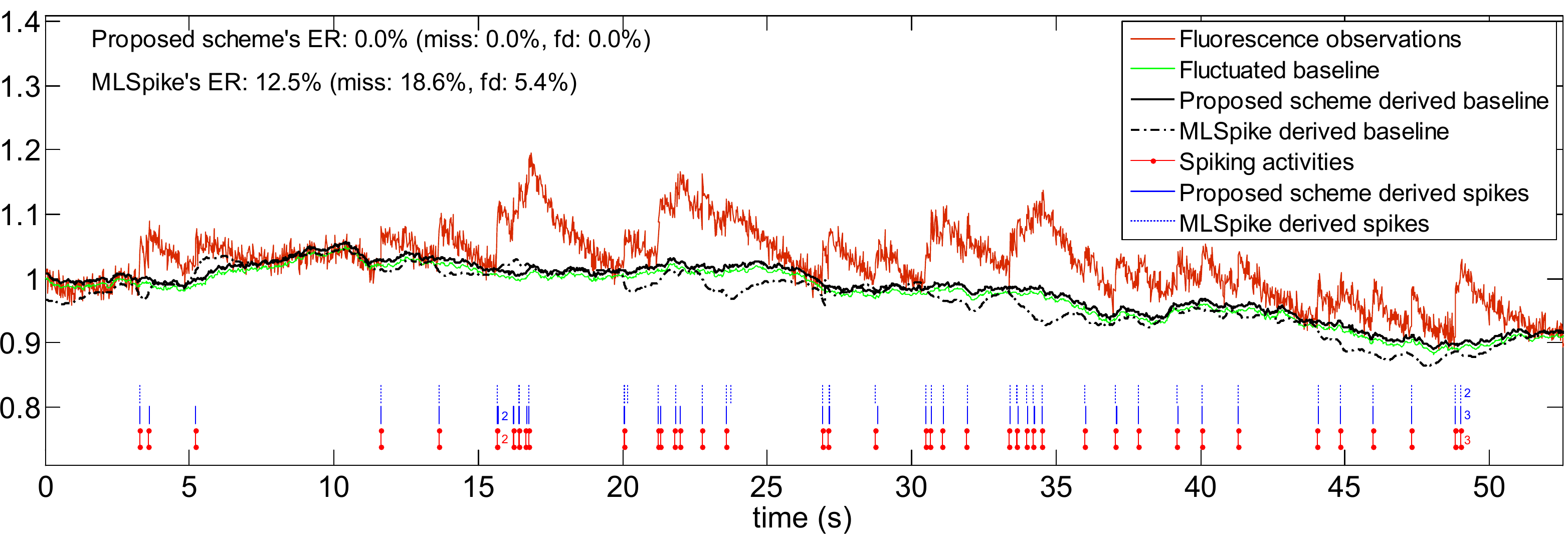}
\caption{One illustration of the performance on estimations of baseline and spikes, with spike rate $\lambda=1$sp/s, and noise level $\alpha=0.2$.}
\end{figure*}

\subsection{Performance on Baseline and Spike Inference}
\begin{figure*}[!t]
\centering
\subfloat[Estimation of baseline.]{\includegraphics[width=2.5in]{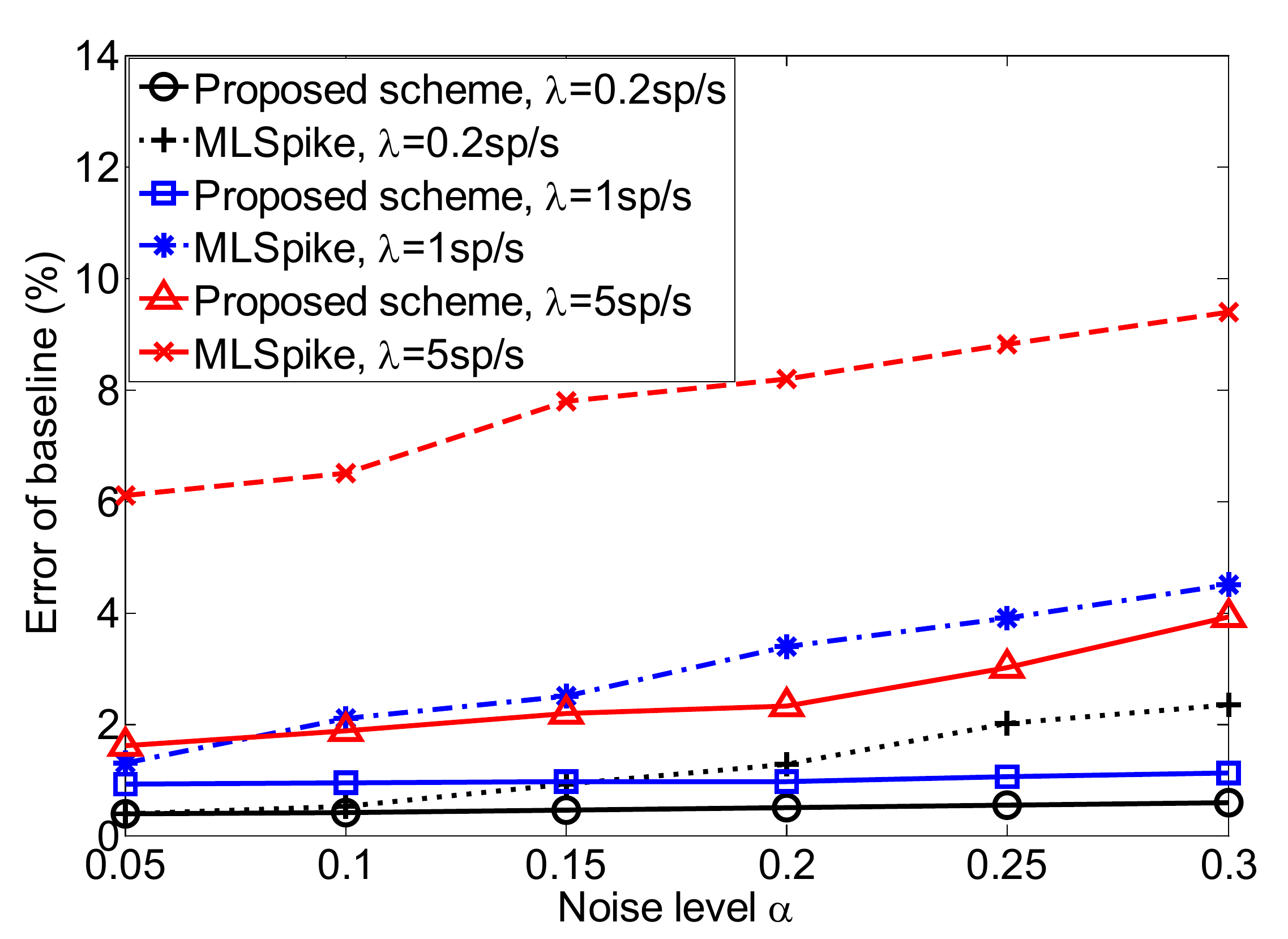}}
\hfil
\subfloat[Detection of informative spikes.]{\includegraphics[width=2.5in]{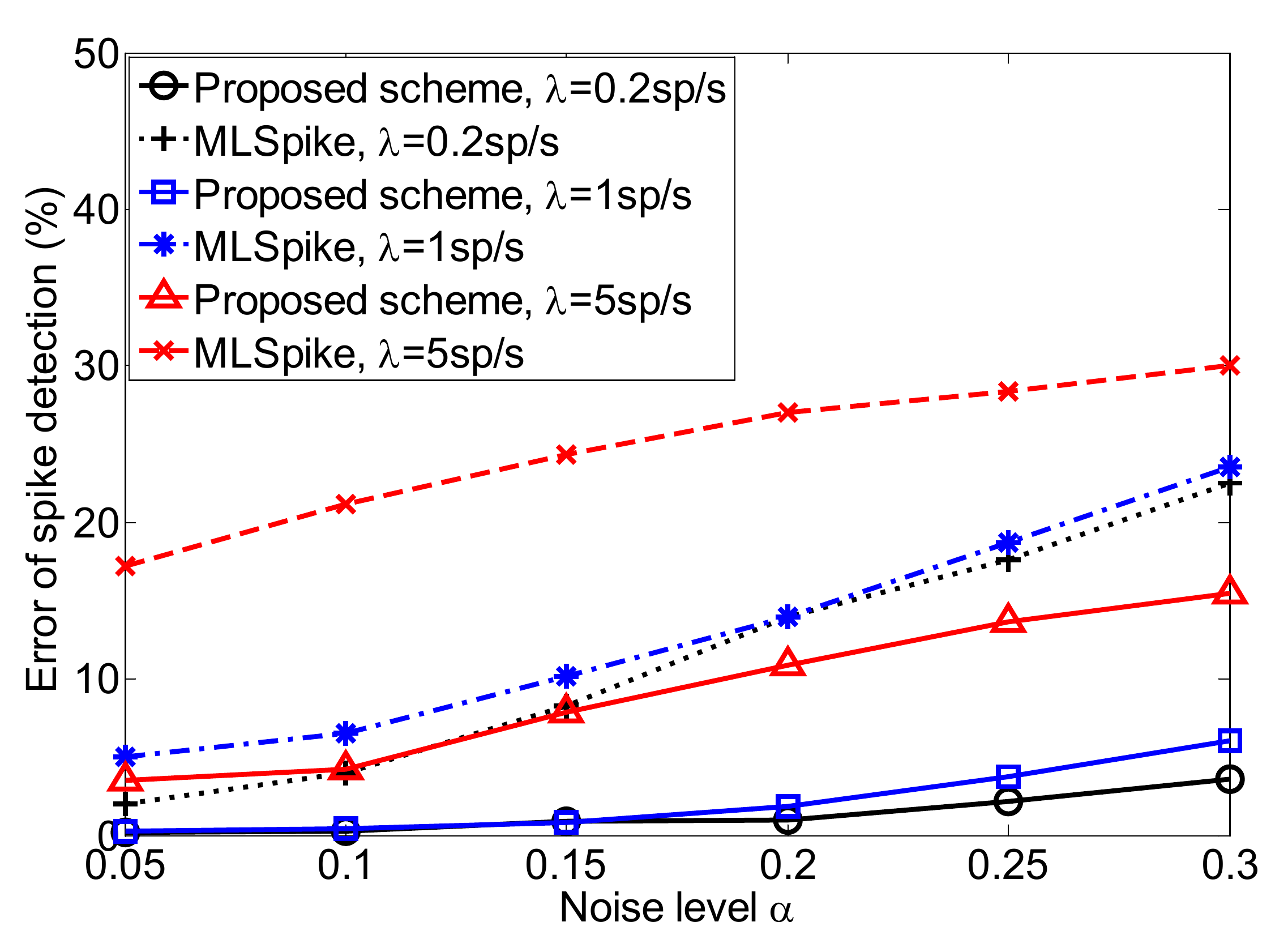}}
\caption{Performance on spike detection and baseline estimation}
\end{figure*}

The performance of the proposed RFS-Bayesian scheme on estimating the fluctuated baseline and detecting the dynamic spikes is provided in Fig. 6 and Fig. 7, in terms of the relative error of baseline estimations, and the average F1-score that reflects the error rate of spike detection.

Typically, the relative error of baseline is defined as follows:
\begin{equation}
\text{RE}_B\triangleq\mathbb{E}\bigg\{\frac{1}{K}\sum_{k=1}^K \frac{||\hat{B}_{k|k}-B_k||_2}{B_k}\bigg\},
\end{equation}
where $K$ denotes for the total number of time-slots. In this experiment, we assign $K=25000$.

The F1-score error rate is the harmonic mean of the \emph{sensitivity} and the \emph{precision}, representing the average of the percentages of misses and false detection \cite{Victor1996Nature}. By respectively providing the sensitivity and the precision as:
$$\text{\emph{sensitivity}}=1-\frac{\text{misses}}{\text{total spikes}},$$
$$\text{\emph{precision}}=1-\frac{\text{false detection}}{\text{total detection}},$$
the F1-score based error rate, denoted as $\epsilon$ can be specified by:
$$\epsilon=1-2\times\frac{\text{\emph{sensitivity}}\times\text{\emph{precision}}} {\text{\emph{sensitivity}}+\text{\emph{precision}}}.$$
Based on the computation of $\epsilon$, the relative error of spike detection is therefore defined as:
\begin{equation}
\text{RE}_s\triangleq\mathbb{E}\{\epsilon\}.
\end{equation}

\subsubsection{Estimation Performance of Baseline}
The $\text{RE}_B$ is shown in the Fig. 7-(a). It is observed that with the rises of both the noise level and the spike rate (which represents increasing interactions by adding more spike activities), $\text{RE}_B$ from both the proposed scheme and the MLSpike scheme is correspondingly growing.

Then, it is noteworthy that the proposed RFS-Bayesian scheme outperforms the MLSpike method. For instance, given a fixed spike rate as $\lambda=0.2$sp/s, the $\text{RE}_B$ from the proposed scheme keeps smaller than that of the MLSpike scheme, especially when noise level is high (i.e., $\alpha>0.2$). The advantages of the proposed RFS-Bayesian is to some extent due to the correct estimations of the aforementioned model parameters, which if biased, may result in erroneous estimation of the baseline.

From the illustration in Fig. 6, we can observe that, for the counterpart MLSpike method, the erroneous derivations of baselines are more likely to happen within the active piece of spikes (e.g., when the time in Fig. 6 is near the $5$s). This is because the MLSpike is unable to estimate the baseline variance, which leads to deviations from the real baseline. By contrast, the proposed RFS-Bayesian is capable of utilizing both the active and silent piece to estimate the baseline variance (seen from Fig. 5-(b)), and thereby has more chances to successfully track the actual baseline fluctuations.

\subsubsection{Detection Performance of Spikes}
The comparison between the RFS-Bayesian scheme and the MLSpike scheme on detecting the dynamic spikes is performed. According to Fig. 7-(b), it can be firstly seen that the performance of both schemes are heavily deteriorated by the increments of the ambient noise and the spike rate, suggesting that the effects of noise and coupling interactions influence the detection performance.

Then, in comparing with the performance of two schemes, we see that the $\text{RE}_s$ from the proposed scheme remains lower than that of the MLSpike scheme. For instance, when $\lambda=1$sp/s, it is illustrated that at a relatively low noise level $\alpha=0.05$, the $\text{RE}_s$ of the proposed scheme approaches less than $1\%$, whilst the value from the MLSpike reaches almost $5\%$. At a high noise level region (e.g., $\alpha=0.3$), the $\text{RE}_s$ value of the MLSpike is greater than $20\%$, but our proposed scheme produces the $\text{RE}_s$ of nearly $5\%$, i.e. the error ratio can be reduced by $75\%$ with our proposed new scheme.

The explanation of the detection advantages from the proposed RFS-Bayesian scheme is given. As shown in Fig. 6, the detection errors from the MLSpike are more likely to occur with erroneous estimations of the baseline (see the $3$-$5$s, and $15$-$25$s from Fig. 6). This is because the erroneous estimation undermines the detection of the correct spikes, which in turn inevitably causes subsequent errors in the next-round baseline/parameter estimations or spike inferences. By contrast, attributed to the unique ability of the RFS-Bayesian scheme to appropriately use the active/silent pieces for inference, the reliability of the estimations on model parameters and baseline can be ensured, which further guarantees the spike detection performance. Besides, the parameter estimation can be continually refined, leading to the more accurate spike transition shape.

\subsection{Complexity Analysis}
\begin{figure}[!t]
\centering
\includegraphics[width=2.5in]{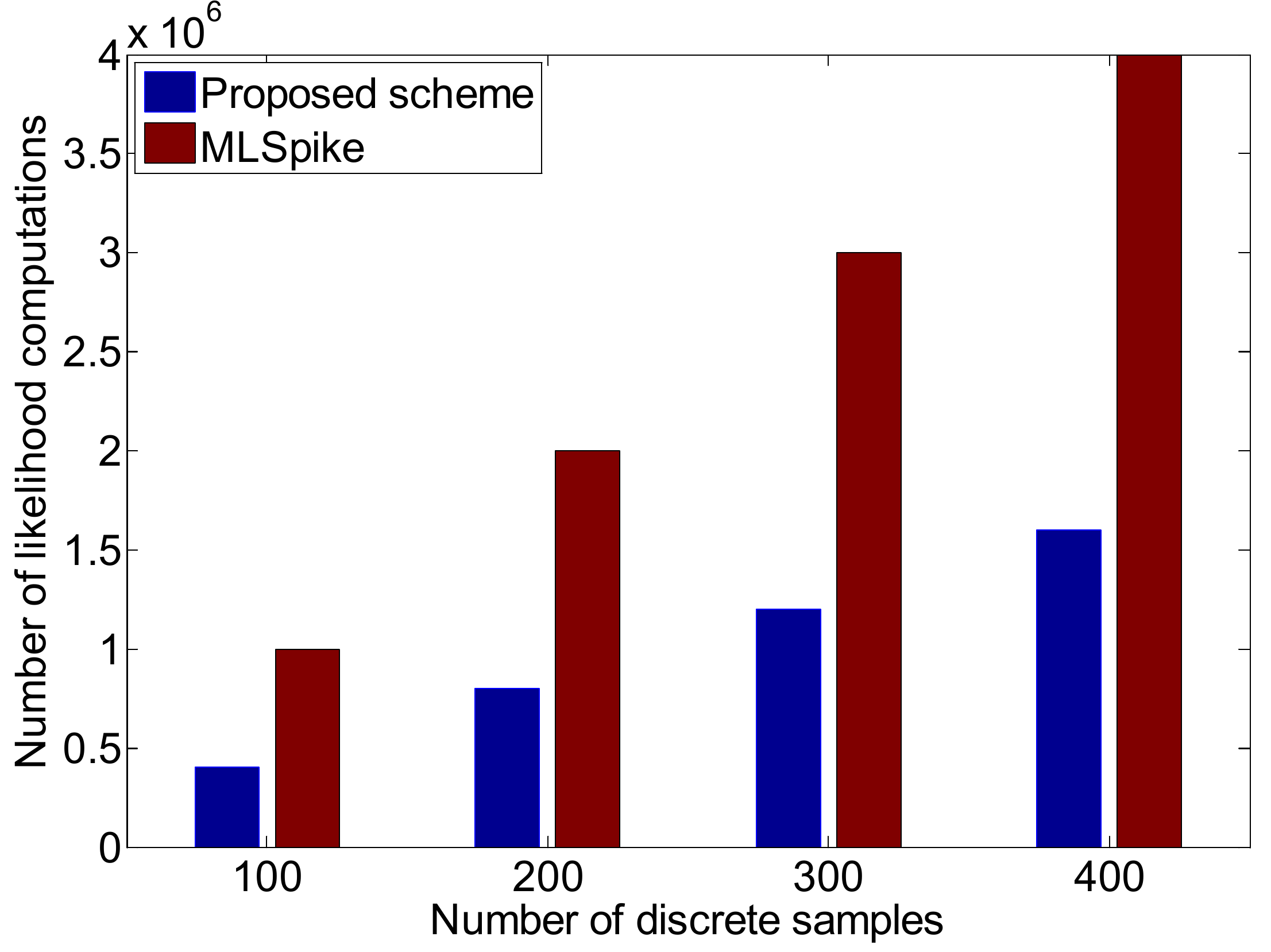}
\caption{Computational complexity comparison}
\end{figure}

We finally evaluate the computational complexity of the proposed RFS-Bayeisan scheme.

In general, the complexity of the Bayesian methods is roughly measured by the total numbers of likelihood computations, each of which has been given as $\vartheta$ \cite{Borwein1998Pi}, which is related with the representative precision and various adopted algorithms. For our proposed RFS-Bayesian scheme, the times of computations are proportional to $2$ times of the size of simulated particles (i.e., $2\times I$) of the PF algorithm, since it computes two likelihoods in Eq. (21)-(22) for each particle. Therefore, the computational complexity of the proposed scheme for each sample is described as $\text{O}\big(2\times I\cdot\vartheta)=2\times2000\times\vartheta$. The total computational complexity with respect to different number of discrete samples is shown in Fig. 8.

By contrast, the complexity of the MLSpike scheme depends on \textbf{i)} the discretion of state space, i.e., $B_k$ and $C_k$, and \textbf{ii)} the enumeration of $s_k$. This is because for each possible assemble (i.e., $[B_k,C_k,s_k]^T$), its related likelihood density should be computed once. In MLSpike scheme given by \cite{Deneux2016Accurate}, $B_k$ and $C_k$ are discretized as a $50\times50$ two-dimension grid, and $s_k$ is enumerated as $0$, $1$, $2$, $3$ (i.e., limit to three spikes per time-slot). As a result, the number of computations on likelihood densities for each sample is $50\times50\times4=10000$, and the complexity is subsequently represented as $10000\times\vartheta$, which is more than two times the proposed RFS-Bayesian scheme. We can see from Fig. 8 that, as the number of samples increases, the difference between the complexities of two schemes grows larger, which if combined with the aforementioned improvement of detection/estimation, suggests that our proposed scheme is capable of reliably inferring the random neuron spikes at the expense of reduced computational complexity.

\section{Conclusion}
In most signal processing tasks of biological or other complex conditions, simultaneously detecting informative signal (i.e. the concerned neuron spikes) and estimating model characteristic is important for scientific research and engineering practice. However, difficulties such as the dynamic of spikes, fluctuated baseline, unknown model parameters, and ambient noise make accurate detection/estimation extremely challenging, which is usually beyond the competence of most existing processing algorithms. In this paper, we have proposed a new RFS-Bayesian scheme, in order to improve the detection/estimation performance in the complicated dynamic environments. By exploiting the different response characteristics of positive spike signal and dynamic background signal, we designed a novel RFS state to uncouple complex interactions and formidable difficulties in jointly recursive estimation and detection. This allows us to structure the challenging signal processing problem into a RFS-based two-stage Bayesian framework. By propagating the RFS state, the detection/estimation can be subsequently solved with increased accuracy compared to the state-of-the-art MLSpike scheme. Our proposed scheme may provide the great promise for future research in adverse biological or engineering applications.

\appendices
\section{Proof of Theorem 1 and Proposition 1}
\emph{Theorem 1} gives the predicted active probability (i.e., $q_{k|k-1}$) and the spatial densities (i.e., $p_{k|k-1}(\mathbf{v}_k)$, and $p_{k|k-1}(\mathbf{u}_k,\mathbf{v}_k)$) by analyzing the predict-stage in Eq. (28).

In the case of $\bm{\mathcal{U}}_k=\{\mathbf{v}_k\}$, Eq. (28) can be addressed via respectively computing the set integral under $\bm{\mathcal{U}}_{k-1}=\{\mathbf{v}_{k-1}\}$, and $\bm{\mathcal{U}}_{k-1}=\{\mathbf{u}_{k-1},\mathbf{v}_{k-1}\}$, i.e.,
\begin{equation}
\begin{aligned}
&f_{k|k-1}(\{\mathbf{v}_k\}|\mathbf{F}_{1:k-1})\\
=&\int\phi_{k|k-1} (\{\mathbf{v}_k\}|\{\mathbf{v}_{k-1}\}) f_{k-1|k-1}(\{\mathbf{v}_{k-1}\}|\mathbf{F}_{1:k-1})d\mathbf{v}_{k-1}\\
&+\iint\phi_{k|k-1} (\{\mathbf{v}_k\}|\{\mathbf{u}_{k-1},\mathbf{v}_{k-1}\})\\ &\cdot f_{k-1|k-1}(\{\mathbf{u}_{k-1},\mathbf{v}_{k-1}\}|\mathbf{F}_{1:k-1}) d\mathbf{u}_{k-1}\mathbf{v}_{k-1},
\end{aligned}
\end{equation}
where $\phi_{k|k-1}(\{\mathbf{v}_k\}|\{\mathbf{v}_{k-1}\})$, $\phi_{k|k-1} (\{\mathbf{v}_k\}|\{\mathbf{u}_{k-1},\mathbf{v}_{k-1}\})$ are the transitional FISST PDFs specified in Eqs. (16)-(17).
It is also noteworthy that $f_{k|k-1}(\{\mathbf{v}_k\}|\mathbf{F}_{1:k-1})$, $f_{k-1|k-1}(\{\mathbf{v}_{k-1}\}|\mathbf{F}_{1:k-1})$, and $f_{k-1|k-1}(\{\mathbf{u}_{k-1},\mathbf{v}_{k-1}\}|\mathbf{F}_{1:k-1})$ are all predicted FISST PDFs, composed of the cardinality probability, and the joint density of the elements. According to Eqs. (23)-(24) we can re-write them as:
\begin{equation}
f_{k|k-1}(\{\mathbf{v}_k\}|\mathbf{F}_{1:k-1})=(1-q_{k|k-1})\cdot p_{k|k-1}(\mathbf{v}_k),
\end{equation}
\begin{equation}
\begin{aligned}
f_{k-1|k-1}(\{\mathbf{v}_{k-1}\}&|\mathbf{F}_{1:k-1})\\
&=(1-q_{k-1|k-1})\cdot p_{k-1|k-1}(\mathbf{v}_{k-1}),
\end{aligned}
\end{equation}
and
\begin{equation}
\begin{aligned}
f_{k-1|k-1}(\{\mathbf{u}_{k-1},&\mathbf{v}_{k-1}\}|\mathbf{F}_{1:k-1})\\
&=(q_{k-1|k-1})\cdot p_{k-1|k-1}(\mathbf{u}_{k-1},\mathbf{v}_{k-1}).
\end{aligned}
\end{equation}

By taking (16)-(17), and (62)-(64) into (61), we derive,
\begin{equation}
\begin{aligned}
&(1-q_{k|k-1})\cdot p_{k|k-1}(\mathbf{v}_k)\\[2mm]
=&(1-q_{k-1|k-1})(1-P_b)\cdot\\
&\int\pi_{k|k-1} (\mathbf{v}_k|\mathbf{v}_{k-1})\cdot p_{k-1|k-1}(\mathbf{v}_{k-1})d\mathbf{v}_{k-1}\\
&+q_{k-1|k-1}\cdot(1-P_s)\cdot\\
&\iint\pi_{k|k-1} (\mathbf{v}_k|\mathbf{v}_{k-1}) p_{k-1|k-1}(\mathbf{u}_{k-1},\mathbf{v}_{k-1})d\mathbf{u}_{k-1}\mathbf{v}_{k-1}\\[2mm]
\overset{(a)}{=}&(1-q_{k-1|k-1})(1-P_b)\cdot\\
&\int\pi_{k|k-1} (\mathbf{v}_k|\mathbf{v}_{k-1})\cdot p_{k-1|k-1}(\mathbf{v}_{k-1})d\mathbf{v}_{k-1}\\
&+q_{k-1|k-1}\cdot(1-P_s)\cdot\\
&\int\pi_{k|k-1} (\mathbf{v}_k|\mathbf{v}_{k-1})\cdot p_{k-1|k-1}(\mathbf{v}_{k-1})d\mathbf{v}_{k-1}\\[2mm]
=&\big[1-P_s\cdot q_{k-1|k-1}-P_b\cdot(1-q_{k-1|k-1})\big]\\
&\cdot\int\pi_{k|k-1} (\mathbf{v}_k|\mathbf{v}_{k-1})\cdot p_{k-1|k-1}(\mathbf{v}_{k-1})d\mathbf{v}_{k-1},
\end{aligned}
\end{equation}
where $(a)$ holds for the fact that $p_{k-1|k-1}(\mathbf{u}_{k-1},\mathbf{v}_{k-1})=p_{k-1|k-1}(\mathbf{u}_{k-1})\cdot p_{k-1|k-1}(\mathbf{v}_{k-1})$, since $\mathbf{u}_{k-1}$ and $\mathbf{v}_{k-1}$ are independent. Recalling the construction of the RFS state $\bm{\mathcal{U}}_k$ where $\mathbf{v}_k$ cannot disappear from the fluorescence observation, $p_{k|k-1}(\mathbf{v}_k)$ therefore can be computed by the famous Chapman-Kolmogorov formula, which proves the Eq. (31) in \emph{Theorem 1}. Then, taking (31) into (65), the predicted \emph{active} probability $q_{k|k-1}$ can be calculated, as specified in Eq. (30).

Next, we will prove Eq. (32) by studying the case $\bm{\mathcal{U}}_k=\{\mathbf{u}_k,\mathbf{v}_k\}$. As aforementioned, Eq. (28) is computed by given $\bm{\mathcal{U}}_{k-1}=\{\mathbf{v}_{k-1}\}$, and $\bm{\mathcal{U}}_{k-1}=\{\mathbf{u}_{k-1},\mathbf{v}_{k-1}\}$, i.e.,
\begin{equation}
\begin{aligned}
&f_{k|k-1}(\{\mathbf{u}_k,\mathbf{v}_k\}|\mathbf{F}_{1:k-1})\\[2mm]
=&\int\phi_{k|k-1} (\{\mathbf{u}_k,\mathbf{v}_k\}|\{\mathbf{v}_{k-1}\})\\ &\cdot f_{k-1|k-1}(\{\mathbf{v}_{k-1}\}|\mathbf{F}_{1:k-1})d\mathbf{v}_{k-1}\\
&+\iint\phi_{k|k-1} (\{\mathbf{u}_k,\mathbf{v}_k\}|\{\mathbf{u}_{k-1},\mathbf{v}_{k-1}\})\\ &\cdot f_{k-1|k-1}(\{\mathbf{u}_{k-1},\mathbf{v}_{k-1}\}|\mathbf{F}_{1:k-1}) d\mathbf{u}_{k-1}\mathbf{v}_{k-1}\\[2mm]
=&\int P_b\cdot\pi_{k|k-1}(\mathbf{v}_k|\mathbf{v}_{k-1})\cdot b_k(\mathbf{u}_k)\\ &\cdot f_{k-1|k-1}(\{\mathbf{v}_{k-1}\}|\mathbf{F}_{1:k-1})d\mathbf{v}_{k-1}\\
&+\iint P_s\cdot\psi_{k|k-1}(\mathbf{u}_k,\mathbf{v}_k|\mathbf{u}_{k-1},\mathbf{v}_{k-1})\\ &\cdot f_{k-1|k-1}(\{\mathbf{u}_{k-1},\mathbf{v}_{k-1}\}|\mathbf{F}_{1:k-1}) d\mathbf{u}_{k-1}\mathbf{v}_{k-1},
\end{aligned}
\end{equation}
where $f_{k|k-1}(\{\mathbf{u}_k,\mathbf{v}_k\}|\mathbf{F}_{1:k-1})$ in the left-hand is the predicted FISST PDF, and can be viewed as the multiplication of the cardinality probability $q_{k|k-1}$, and the joint distribution $p_{k|k-1}(\mathbf{u}_k,\mathbf{v}_k)$, i.e.,
\begin{equation}
f_{k|k-1}(\{\mathbf{u}_k,\mathbf{v}_k\}|\mathbf{F}_{1:k-1})=q_{k|k-1}\cdot p_{k|k-1}(\mathbf{u}_k,\mathbf{v}_k).
\end{equation}

Taking (63)-(64), and (67) into (66), we have,
\begin{equation}
\begin{aligned}
&q_{k|k-1}\cdot p_{k|k-1}(\mathbf{u}_k,\mathbf{v}_k)\\[2mm]
=&P_b\cdot(1-q_{k-1|k-1})\cdot b_k(\mathbf{u}_k)\\
&\cdot\int\pi_{k|k-1}(\mathbf{v}_k|\mathbf{v}_{k-1})\cdot p_{k-1|k-1}(\mathbf{v}_{k-1})d\mathbf{v}_{k-1}\\
&+P_s\cdot q_{k-1|k-1}\cdot\iint \psi_{k|k-1}(\mathbf{u}_k,\mathbf{v}_k|\mathbf{u}_{k-1},\mathbf{v}_{k-1})\\
&\cdot p_{k-1|k-1}(\mathbf{u}_{k-1},\mathbf{v}_{k-1}) d\mathbf{u}_{k-1}\mathbf{v}_{k-1}\\[2mm]
=&P_b\cdot(1-q_{k-1|k-1})\cdot b_k(\mathbf{u}_k)\cdot p_{k|k-1}(\mathbf{v}_k)\\
&+P_s\cdot q_{k-1|k-1}\cdot\iint \psi_{k|k-1}(\mathbf{u}_k,\mathbf{v}_k|\mathbf{u}_{k-1},\mathbf{v}_{k-1})\\
&\cdot p_{k-1|k-1}(\mathbf{u}_{k-1},\mathbf{v}_{k-1}) d\mathbf{u}_{k-1}\mathbf{v}_{k-1}.
\end{aligned}
\end{equation}
By dividing $q_{k|k-1}$ in the both sides of the (68), Eq. (32) is derived, and thereby the \emph{Theorem 1} is completely proved.

Note from Eq. (68) that the right-hand can be subsequently processed, by respectively addressing the integral as:
\begin{equation}
\begin{aligned}
&\iint\psi_{k|k-1}(\mathbf{u}_k,\mathbf{v}_k|\mathbf{u}_{k-1},\mathbf{v}_{k-1})\\
&\cdot p_{k-1|k-1}(\mathbf{u}_{k-1},\mathbf{v}_{k-1}) d\mathbf{u}_{k-1}\mathbf{v}_{k-1}\\
=&\int\pi_{k|k-1}(\mathbf{v}_{k}|\mathbf{v}_{k-1})\cdot p_{k-1|k-1}(\mathbf{v}_{k-1})d\mathbf{v}_{k-1}\\
&\cdot\int\varpi_{k|k-1}(\mathbf{u}_{k}|\mathbf{u}_{k-1})\cdot p_{k-1|k-1}(\mathbf{u}_{k-1})d\mathbf{u}_{k-1}\\
=&p_{k|k-1}(\mathbf{v}_{k-1}) \int\varpi_{k|k-1}(\mathbf{u}_{k}|\mathbf{u}_{k-1}) p_{k-1|k-1}(\mathbf{u}_{k-1})d\mathbf{u}_{k-1}.
\end{aligned}
\end{equation}
By introducing (69) into (68), Eqs. (33)-(34) are obtained, and thus the \emph{Proposition 1} is proved.

\section{Proof of Theorem 2}
The results of \emph{Theorem 2} specify the posterior active probability (i.e., $q_{k|k}$), and the spatial densities (i.e., $p_{k|k}(\mathbf{u}_k,\mathbf{v}_k)$, and $p_{k|k}(\mathbf{v}_k)$), which are derived from the update-stage in Eq. (29). Note that the denominator in (29) is a normalizing parameter, so we here give a convenient form of (29) as follows:
\begin{equation}
f_{k|k}(\bm{\mathcal{U}}_k|\mathbf{F}_{1:k})\propto\varphi_k(F_k|\bm{\mathcal{U}}_k) \cdot f_{k|k-1}(\bm{\mathcal{U}}_k|\mathbf{F}_{1:k-1}).
\end{equation}
Then, the analysis is equivalent as studying the two cases of the RFS state $\bm{\mathcal{U}}_k$, i.e., $\bm{\mathcal{U}}_k=\{\mathbf{u}_k,\mathbf{v}_k\}$, and $\bm{\mathcal{U}}_k=\{\mathbf{v}_k\}$.

In the case $\bm{\mathcal{U}}_k=\{\mathbf{u}_k,\mathbf{v}_k\}$, where the fluorescence observation $F_k$ involves the information of both $\mathbf{u}_k$, and $\mathbf{v}_k$, the likelihood PDF is $\varphi_k(F_k|\bm{\mathcal{U}}_k)=\varphi_k(F_k|\{\mathbf{u}_k,\mathbf{v}_k\})$. In this view, Eq. (70) can be re-written by replacing $f_{k|k}(\bm{\mathcal{U}}_k|\mathbf{F}_{1:k})=q_{k|k}\cdot p_{k|k}(\mathbf{u}_k,\mathbf{v}_k)$, and $f_{k|k-1}(\bm{\mathcal{U}}_k|\mathbf{F}_{1:k-1})=q_{k|k-1}\cdot p_{k|k-1}(\mathbf{u}_k,\mathbf{v}_k)$, i.e.,
\begin{equation}
\begin{aligned}
q_{k|k}\cdot p_{k|k}&(\mathbf{u}_k,\mathbf{v}_k)\\ &\propto q_{k|k-1}\cdot\varphi_k(F_k|\{\mathbf{u}_k,\mathbf{v}_k\})\cdot p_{k|k-1}(\mathbf{u}_k,\mathbf{v}_k).
\end{aligned}
\end{equation}
Given the definition of $q_{k|k}$, we can update it via computing the portion specified by the likelihood PDF $\varphi_k(F_k|\{\mathbf{u}_k,\mathbf{v}_k\})$, i.e.,
\begin{equation}
q_{k|k}\propto q_{k|k-1}\iint\varphi_k(F_k|\{\mathbf{u}_k,\mathbf{v}_k\}) p_{k|k-1}(\mathbf{u}_k,\mathbf{v}_k)d\mathbf{u}_k\mathbf{v}_k,
\end{equation}
which gives the proof of Eq. (35). By subsequently taking (72) into (71), we have, \begin{equation}
p_{k|k}(\mathbf{u}_k,\mathbf{v}_k) \propto\varphi_k(F_k|\{\mathbf{u}_k,\mathbf{v}_k\})\cdot p_{k|k-1}(\mathbf{u}_k,\mathbf{v}_k),
\end{equation}
which proves the Eq. (36). The posterior spatial PDF $p_{k|k}(\mathbf{v}_k)$ in Eq. (37) is thereby computed from the margin distribution of $p_{k|k}(\mathbf{u}_k,\mathbf{v}_k)$.

Consider the case where $\bm{\mathcal{U}}_k=\{\mathbf{v}_k\}$, i.e., the observation $F_k$ reflects only the state of $\mathbf{v}_k$. In this case, Eq. (70) is viewed as:
\begin{equation}
\begin{aligned}
(1-q_{k|k})\cdot&p_{k|k}(\mathbf{v}_k)\\
&\propto(1-q_{k|k-1})\cdot\varphi_k(F_k|\{\mathbf{v}_k\})\cdot p_{k|k-1}(\mathbf{v}_k).
\end{aligned}
\end{equation}
According to (72), $1-q_{k|k}$ in (74) can be computed as
\begin{equation}
1-q_{k|k}\propto (1-q_{k|k-1})\int\varphi_k(F_k|\{\mathbf{v}_k\}) p_{k|k-1}(\mathbf{v}_k)d\mathbf{v}_k.
\end{equation}
Therefore, taking (75) into (74), the spatial $p_{k|k}(\mathbf{v}_k)$ is derived, i.e.,
\begin{equation}
p_{k|k}(\mathbf{v}_k) \propto\varphi_k(F_k|\{\mathbf{v}_k\})\cdot p_{k|k-1}(\mathbf{v}_k).
\end{equation}
It is also noteworthy that in the case $\bm{\mathcal{U}}_k=\{\mathbf{v}_k\}$, the joint spatial density equals to its marginal distribution, i.e., $p_{k|k-1}(\mathbf{u}_k,\mathbf{v}_k)=p_{k|k-1}(\mathbf{v}_k)$; and the likelihood PDF $\varphi_k(F_k|\{\mathbf{u}_k,\mathbf{v}_k\})$ degrades to $\varphi_k(F_k|\{\mathbf{v}_k\})$. Therefore, Eq. (76) equals to the derivation from the margin distribution of $p_{k|k}(\mathbf{u}_k,\mathbf{v}_k)$, which accomplishes the proof of \emph{Theorem 2}.

\bibliographystyle{IEEEtran}
\bibliography{myref}

\begin{thebibliography}{10}
\providecommand{\url}[1]{#1}
\csname url@samestyle\endcsname
\providecommand{\newblock}{\relax}
\providecommand{\bibinfo}[2]{#2}
\providecommand{\BIBentrySTDinterwordspacing}{\spaceskip=0pt\relax}
\providecommand{\BIBentryALTinterwordstretchfactor}{4}
\providecommand{\BIBentryALTinterwordspacing}{\spaceskip=\fontdimen2\font plus
\BIBentryALTinterwordstretchfactor\fontdimen3\font minus
  \fontdimen4\font\relax}
\providecommand{\BIBforeignlanguage}[2]{{%
\expandafter\ifx\csname l@#1\endcsname\relax
\typeout{** WARNING: IEEEtran.bst: No hyphenation pattern has been}%
\typeout{** loaded for the language `#1'. Using the pattern for}%
\typeout{** the default language instead.}%
\else
\language=\csname l@#1\endcsname
\fi
#2}}
\providecommand{\BIBdecl}{\relax}
\BIBdecl

\bibitem{6112225}
M.~Temerinac-Ott, O.~Ronneberger, P.~Ochs, W.~Driever, T.~Brox, and
  H.~Burkhardt, ``Multiview deblurring for 3-d images from light-sheet-based
  fluorescence microscopy,'' \emph{IEEE Transactions on Image Processing},
  vol.~21, no.~4, pp. 1863--1873, 2012.

\bibitem{Denk73}
W.~Denk, J.~Strickler, and W.~Webb, ``Two-photon laser scanning fluorescence
  microscopy,'' \emph{Science}, vol. 248, no. 4951, pp. 73--76, 1990.

\bibitem{Zipfel2003Nonlinear}
W.~R. Zipfel, R.~M. Williams, and W.~W. Webb, ``Nonlinear magic: multiphoton
  microscopy in the biosciences,'' \emph{Nature Biotechnology}, vol.~21,
  no.~11, pp. 1369--1377, 2003.

\bibitem{Katona2012Fast}
G.~e.~a. Katona, ``Fast two-photon in vivo imaging with three-dimensional
  random-access scanning in large tissue volumes.'' \emph{Nature Methods},
  vol.~9, no.~2, pp. 201--208, 2012.

\bibitem{6708551}
D.~Kilinc and O.~B. Akan, ``Receiver design for molecular communication,''
  \emph{IEEE Journal on Selected Areas in Communications}, vol.~31, no.~12, pp.
  705--714, 2013.

\bibitem{Grewe2010High}
B.~F. Grewe, D.~Langer, H.~Kasper, B.~M. Kampa, and F.~Helmchen, ``High-speed
  in vivo calcium imaging reveals neuronal network activity with
  near-millisecond precision,'' \emph{Nature Methods}, vol.~7, no.~6, pp.
  399--405, 2010.

\bibitem{Chen2013Ultrasensitive}
T.~W. Chen, T.~J. Wardill, Y.~Sun, S.~R. Pulver, S.~L. Renninger, A.~Baohan,
  E.~R. Schreiter, R.~A. Kerr, M.~B. Orger, and V.~Jayaraman, ``Ultrasensitive
  fluorescent proteins for imaging neuronal activity.'' \emph{Nature}, vol.
  499, no. 7458, pp. 295--300, 2013.

\bibitem{Dana2014Thy1}
H.~Dana, T.~W. Chen, A.~Hu, B.~C. Shields, C.~Guo, L.~L. Looger, D.~S. Kim, and
  K.~Svoboda, ``Thy1-gcamp6 transgenic mice for neuronal population imaging in
  vivo,'' \emph{Plos One}, vol.~9, no.~9, p. e108697, 2014.

\bibitem{Akerboom2012Optimization}
J.~Akerboom, T.~W. Chen, T.~J. Wardill, L.~Tian, J.~S. Marvin, S.~Mutlu, N.~C.
  Calderson, F.~Esposti, B.~G. Borghuis, and X.~R. Sun, ``Optimization of a
  gcamp calcium indicator for neural activity imaging.'' \emph{Journal of
  Neuroscience the Official Journal of the Society for Neuroscience}, vol.~32,
  no.~40, pp. 13\,819--40, 2012.

\bibitem{Deneux2016Accurate}
T.~Deneux, A.~Kaszas, G.~Szalay, G.~Katona, T.~Lakner, A.~Grinvald,
  B.~R\'{o}zsa, and I.~Vanzetta, ``Accurate spike estimation from noisy calcium
  signals for ultrafast three-dimensional imaging of large neuronal populations
  in vivo,'' \emph{Nature Communications}, vol.~7, p. 12190, 2016.

\bibitem{Vogelstein2009Spike}
J.~T. Vogelstein and A.~M. Watson~BOPacker, ``Spike inference from calcium
  imaging using sequential monte carlo methods,'' \emph{Biophysical Journal},
  vol.~97, no.~2, p. 636, 2009.

\bibitem{Henry2013Inference}
L.~Henry, G.~Felipe, Z.~Friedemann, G.~Wulfram, and H.~Fritjof, ``Inference of
  neuronal network spike dynamics and topology from calcium imaging data,''
  \emph{Frontiers in Neural Circuits}, vol.~7, no.~7, p. 201, 2013.

\bibitem{Kerr2005From}
J.~N. Kerr, D.~Greenberg, and F.~Helmchen, ``From the cover: Imaging input and
  output of neocortical networks in vivo,'' \emph{Proc Natl Acad Sci U S A},
  vol. 102, no.~39, pp. 14\,063--14\,068, 2005.

\bibitem{Kerr2007Spatial}
J.~N. Kerr, C.~P. de~Kock, D.~S. Greenberg, R.~M. Bruno, B.~Sakmann, and
  F.~Helmchen, ``Spatial organization of neuronal population responses in layer
  2/3 of rat barrel cortex.'' \emph{Journal of Neuroscience the Official
  Journal of the Society for Neuroscience}, vol.~27, no.~48, pp.
  13\,316--13\,328, 2007.

\bibitem{Greenberg2008Population}
D.~S. Greenberg, A.~R. Houweling, and J.~N. Kerr, ``Population imaging of
  ongoing neuronal activity in the visual cortex of awake rats.'' \emph{Nature
  Neuroscience}, vol.~11, no.~7, pp. 749--751, 2008.

\bibitem{Ozden2008}
I.~Ozden, H.~M. Lee, M.~R. Sullivan, and S.~S.-H. Wang, ``Identification and
  clustering of event patterns from in vivo multiphoton optical recordings of
  neuronal ensembles,'' \emph{Journal of Neurophysiology}, vol. 100, no.~1, pp.
  495--503, 2008.

\bibitem{Ranganathan2010Optical}
G.~N. Ranganathan and H.~J. Koester, ``Optical recording of neuronal spiking
  activity from unbiased populations of neurons with high spike detection
  efficiency and high temporal precision.'' \emph{Journal of Neurophysiology},
  vol. 104, no.~3, pp. 1812--24, 2010.

\bibitem{8309355}
S.~Soltanian-Zadeh, Y.~Gong, and S.~Farsiu, ``Information-theoretic approach
  and fundamental limits of resolving two closely-timed neuronal spikes in
  mouse brain calcium imaging,'' \emph{IEEE Transactions on Biomedical
  Engineering}, 2018.

\bibitem{6810293}
E.~A. Pnevmatikakis, J.~Merel, A.~Pakman, and L.~Paninski, ``Bayesian spike
  inference from calcium imaging data,'' in \emph{2013 Asilomar Conference on
  Signals, Systems and Computers}, 2013, pp. 349--353.

\bibitem{Pnevmatikakis2016Simultaneous}
E.~Pnevmatikakis, D.~Soudry, Y.~Gao, T.~A. Machado, J.~Merel, D.~Pfau,
  T.~Reardon, M.~Yu, C.~Lacefield, and W.~Yang, ``Simultaneous denoising,
  deconvolution, and demixing of calcium imaging data,'' \emph{Neuron},
  vol.~89, no.~2, pp. 285--299, 2016.

\bibitem{6497685}
B.~Ristic, B.~T. Vo, B.~N. Vo, and A.~Farina, ``A tutorial on bernoulli
  filters: Theory, implementation and applications,'' \emph{IEEE Transactions
  on Signal Processing}, vol.~61, no.~13, pp. 3406--3430, 2013.

\bibitem{Mahler2007Statistical}
R.~P.~S. Mahler, \emph{Statistical Multisource-Multitarget Information
  Fusion}.\hskip 1em plus 0.5em minus 0.4em\relax Artech House, Inc., 2007.

\bibitem{7423817}
B.~Li, J.~Hou, X.~Li, Y.~Nan, A.~Nallanathan, and C.~Zhao, ``Deep sensing for
  space-time doubly selective channels: When a primary user is mobile and the
  channel is flat rayleigh fading,'' \emph{IEEE Transactions on Signal
  Processing}, vol.~64, no.~13, pp. 3362--3375, 2016.

\bibitem{1236770}
P.~M. Djuric, J.~H. Kotecha, J.~Zhang, Y.~Huang, T.~Ghirmai, M.~F. Bugallo, and
  J.~Miguez, ``Particle filtering,'' \emph{IEEE Signal Processing Magazine},
  vol.~20, no.~5, pp. 19--38, 2003.

\bibitem{Victor1996Nature}
J.~D. Victor and K.~P. Purpura, ``Nature and precision of temporal coding in
  visual cortex: a metric-space analysis.'' \emph{Journal of Neurophysiology},
  vol.~76, no.~2, pp. 1310--26, 1996.

\bibitem{Borwein1998Pi}
J.~M. Borwein and P.~B. Borwein, ``Pi and the agm: a study in the analytic
  number theory and computational complexity,'' \emph{Mathematics of
  Computation}, vol.~50, no. 181, pp. 723--729, 1998.

\end{thebibliography}

\end{document}